\documentclass[twocolumn]{aastex6}
\usepackage{multirow, graphicx, epsfig}
\usepackage{soul}
\usepackage{ulem} 
\usepackage{amsmath}
\DeclareGraphicsExtensions{.eps}
\usepackage{wasysym}
\usepackage{subfigure}

\usepackage[nolist]{acronym} 
\usepackage{filecontents} 

\slugcomment{Draft version \today}
\shorttitle{Modeling the SEDs and Spectropolarimetry of Blazars -- 4C+01.02 in 2016--2017}
\shortauthors{Schutte et al.}
\newcommand{\MBH}{\sim 3 \times 10^{9}}
\newcommand{\Flarenz}{3.3 \times 10^{49} }
\newcommand{\Flaregoe}{ 54.8}
\newcommand{\Flaregbe}{7.27 \times 10^2}
\newcommand{\Flaregce}{3.00 \times 10^3}
\newcommand{\Flarevmo}{5.0\times 10^{10} }
\newcommand{\Flarevbo}{8.8\times 10^{12} }
\newcommand{\Flarevco}{1.5\times 10^{14}}
\newcommand{\Flarepo}{2.62}
\newcommand{\Flarept}{3.00}
\newcommand{\Flarefb}{0.188}
\newcommand{\Flareld}{4.5 \times 10^{46}}
\newcommand{\Flaretm}{3.5 \times 10^{4} } 
\newcommand{\Flareho}{4.8 \times 10^{10}}
\newcommand{\Flareht}{1.0 \times 10^{11}}
\newcommand{\Flarehth}{3.1 \times 10^{10}}
\newcommand{\Flareao}{ 0.81} 
\newcommand{\Flareat}{ 1.00}
\newcommand{\Flareso}{0.73} 
\newcommand{\Flarest}{0.75}
\newcommand{\Flarechi}{2.88}
\newcommand{\Flarevmio}{6.6 \times 10^{14}}
\newcommand{\Quiesnz}{1.1 \times 10^{50}}
\newcommand{\Quiesgoe}{24.5}
\newcommand{\Quiesgbe}{4.90 \times 10^{2}}
\newcommand{\Quiesgce}{1.51 \times 10^{3}}
\newcommand{\Quiesvmo}{1.0 \times 10^{10}}
\newcommand{\Quiesvbo}{4.0 \times 10^{12}}
\newcommand{\Quiesvco}{3.8 \times 10^{13}}
\newcommand{\Quiespo}{2.60}
\newcommand{\Quiespt}{3.00}
\newcommand{\Quiesfb}{0.040}
\newcommand{\Quiesld}{3.7 \times 10^{46} }
\newcommand{\Quiestm}{ 3.4 \times 10^{4}}
\newcommand{\Quiesho}{ 4.8 \times 10^{10}}
\newcommand{\Quiesht}{1.0 \times 10^{11}}
\newcommand{\Quieshth}{3.1 \times 10^{10}}
\newcommand{\Quiesao}{0.80} 
\newcommand{\Quiesat}{1.00}
\newcommand{\Quiesso}{ 0.73} 
\newcommand{\Quiesst}{0.75}
\newcommand{\Quieschi}{1.46}
\newcommand{\Quiesvmio}{6.3 \times 10^{14}}
\newcommand{\Flarekelum}{3.2 \times 10^{45}}
\newcommand{\Quieskelum}{6.0 \times 10^{45}}
\newcommand{\Flareb}{0.82}
\newcommand{\Quiesb}{0.82}
\newcommand{\Flareemh}{0.15}
\newcommand{\Quiesemh}{0.30}
\newcommand{\FlareGamma}{15}
\newcommand{\QuiesGamma}{15}
\newcommand{\Flarebr}{3.0 \times 10^{17}}
\newcommand{\Quiesbr}{3.0\times 10^{17}}
\newcommand{\Flareobsa}{3.5} 
\newcommand{\Quiesobsa}{3.5} 
\newcommand{\Flareu}{9.0\times 10^{-3}}
\newcommand{\Quiesu}{5.5\times 10^{-4}} 
\newcommand{\Flarebbt}{5\times 10^{4}} 
\newcommand{\Quiesbbt}{5\times 10^{4}} 
\newcommand{\Flarelb}{5.1 \times 10^{46}} 
\newcommand{\Quieslb}{5.1 \times 10^{46}}
\newcommand{\Flarelble}{16.0}
\newcommand{\Quieslble}{8.51}
\newcommand{\Flaredtvars}{1.9\times 10^{6}} 
\newcommand{\Quiesdtvars}{1.9 \times 10^{6}}
\newcommand{\Flaredtvarh}{5.3\times 10^{2}} 
\newcommand{\Quiesdtvarh}{5.3 \times 10^{2}} 
\newcommand{\GhiMBH}{\sim 5 \times 10^{9} }
\newcommand{\FlarePSb}{0.82}
\newcommand{\QuiesPSb}{0.82}
\newcommand{\FlarePSGamma}{15}
\newcommand{\QuiesPSGamma}{15}
\newcommand{\FlarenzPS}{3.9 \times 10^{49}}
\newcommand{\FlaregoePS}{54.8}
\newcommand{\FlaregbePS}{7.26 \times 10^2}
\newcommand{\FlaregcePS}{2.45 \times 10^3}
\newcommand{\FlarevmoPS}{5.0 \times 10^{10} }
\newcommand{\FlarevboPS}{8.8 \times 10^{12} }
\newcommand{\FlarevcoPS}{1.0 \times 10^{14}}
\newcommand{\FlarepoPS}{2.50}
\newcommand{\FlareptPS}{3.00}
\newcommand{\FlarefbPS}{0.21}
\newcommand{\FlareldPS}{5.2 \times 10^{46}}
\newcommand{\FlaretmPS}{2.8 \times 10^{4} } 
\newcommand{\FlarehoPS}{4.8 \times 10^{10}}
\newcommand{\FlarehtPS}{1.0 \times
 10^{11}}
\newcommand{\FlarehthPS}{3.1 \times 10^{10}}
\newcommand{\FlareaoPS}{0.75} 
\newcommand{\FlareatPS}{1.00}
\newcommand{\FlaresoPS}{0.72} 
\newcommand{\FlarestPS}{0.75}
\newcommand{\FlarechiPS}{3.15}
\newcommand{\FlarevmioPS}{5.3 \times 10^{14}}
\newcommand{\QuiesnzPS}{1.3 \times 10^{50}}
\newcommand{\QuiesgoePS}{24.5}
\newcommand{\QuiesgbePS}{4.88 \times 10^{2}}
\newcommand{\QuiesgcePS}{1.24 \times 10^{3}}
\newcommand{\QuiesvmoPS}{1.0 \times 10^{9} }
\newcommand{\QuiesvboPS}{4.0 \times 10^{12}}
\newcommand{\QuiesvcoPS}{2.5 \times 10^{13} }
\newcommand{\QuiespoPS}{2.33}
\newcommand{\QuiesptPS}{3.00}
\newcommand{\QuiesfbPS}{0.053}
\newcommand{\QuiesldPS}{4.4 \times 10^{46}}
\newcommand{\QuiestmPS}{2.7 \times 10^{4} }
\newcommand{\QuieshoPS}{4.8 \times 10^{10}}
\newcommand{\QuieshtPS}{1.0 \times
 10^{11}}
\newcommand{\QuieshthPS}{3.1 \times 10^{10}}
\newcommand{\QuiesaoPS}{0.66}
\newcommand{\QuiesatPS}{1.00}
\newcommand{\QuiessoPS}{0.71} 
\newcommand{\QuiesstPS}{0.75}
\newcommand{\QuieschiPS}{1.53}
\newcommand{\QuiesvmioPS}{5.1 \times 10^{14}}

\usepackage{tikz,xcolor,hyperref}
\definecolor{lime}{HTML}{A6CE39}
\DeclareRobustCommand{\orcidicon}{%
	\begin{tikzpicture}
	\draw[lime, fill=lime] (0,0) 
	circle [radius=0.16] 
	node[white] {{\fontfamily{qag}\selectfont {\mbox{}\newline \tiny \textbf{iD}}}};
	\draw[white, fill=white]  (-0.0625,0.05)
	circle [radius=0.008];
	\end{tikzpicture}
	\hspace{-3.75mm} 
}
\foreach \x in {A, ..., Z}{%
	\expandafter\xdef\csname orcid\x\endcsname{\noexpand\href{https://orcid.org/\csname orcidauthor\x\endcsname}{\noexpand\orcidicon}}
}

\begin{document}
\title{Modeling the Spectral Energy Distributions and Spectropolarimetry of Blazars -- Application to 4C+01.02 in 2016--2017$^\star$}\thanks{$^\star$based on observations made with the Southern African Large Telescope (SALT) under programme 2016-2-LSP-001 (PI: D.~A.~H. Buckley)}
\author{Hester M.~Schutte$^{1, \, \dagger}$\orcidA{}, Richard J.~Britto$^{2, \, \dagger\dagger}$\orcidB{}, Markus~B\"{o}ttcher$^1$\orcidC{}, Brian~van Soelen$^2$\orcidD{}, \mbox{Johannes P.~Marais$^2$\orcidE{}}, Amanpreet~Kaur$^4$\orcidF{}, Abraham D.~Falcone$^4$\orcidG{}, David A.H.~Buckley$^{2, \, 3}$\orcidH{}, Andry~F.~Rajoelimanana$^2$\orcidI{} and Justin Cooper$^2$\orcidJ{}}
\email{Corresponding authors:\\ $^\dagger$ schuttehester1@nwu.ac.za,\\ $^{\dagger\dagger}$ dr.richard.britto@gmail.com}
\affil{$^1$Centre for Space Research, North-West University, Potchefstroom 2520, South Africa\\ $^2$Department of Physics, University of the Free State, Bloemfontein 9300, South Africa\\ $^3$South African Astronomical Observatory, Observatory, Cape Town 7935, South Africa\\ $^4$Department of Astronomy and Astrophysics, Pennsylvania State University, University Park, PA 16802, United States of America\\ \\(Accepted for publication in ApJ)}

\begin{abstract} 
The optical radiation emitted by blazars contains contributions from synchrotron radiation by relativistic electrons in the jets, as well as thermal radiation emitted mainly by the \ac{AD}, the \ac{BLR} and the host galaxy. The unpolarized radiation components from the \ac{AD}, \ac{BLR} and host galaxy present themselves by decreasing the total polarization in the optical/\ac{UV} spectrum.
A combined model for the \ac{SED} and degree of optical/\ac{UV} polarization is constructed, enabling the disentanglement of the synchrotron and \ac{AD} components.
Our model is applied to the multi-wavelength \ac{SED} and spectropolarimetry observations of the Flat Spectrum Radio Quasar 4C+01.02 ($z=2.1$) in its 2016 July-August flaring state and July-August 2017 quiescent state, using data from the \emph{Fermi} Large Area Telescope, the \textit{Southern African Large Telescope} and the \textit{Las Cumbres Observatory} network of telescopes. By constraining the \ac{AD} component, the mass of the \ac{SMBH} is obtained as $\MBH \, \rm M_{\astrosun}$. Furthermore, the model retrieves the characteristics of the relativistic electron distribution in the jet and the degree of ordering of the magnetic field. Our results highlight the potential of spectropolarimetry observations for disentangling thermal from non-thermal (jet) emission components and thus revealing the physics of particle acceleration and high-energy emission in active galactic nuclei jets.
\end{abstract}


\acresetall  

\section{Introduction}\label{sec:Introduction}
Blazars are a class of jet-dominated (radio-loud) \ac{AGN} in which one of the jets is closely aligned to our line of sight, leading to strongly Doppler boosted emission received by the observer. They can be subdivided into two classes, namely, \ac{FSRQ} and \mbox{BL Lac} objects, which are distinguishable by the visibility of their emission line features in optical spectra: \ac{FSRQ}s have broad emission lines (equivalent width $>5$ \AA{}) while \mbox{BL Lacs} have weak or absent  emission lines \citep{1991ApJ...374..431S}. Blazars are characterized by rapid variability across the electromagnetic spectrum and a high degree of polarization in the radio and optical regime. 

The detection of significantly polarized optical emission from many blazars indicates that it originates dominantly from synchrotron radiation \citep{1986rpa..book.....R, 1986ApJ...305..484S}. Additional, unpolarized emission components arising from the dust torus (Infrared, IR), host galaxy (optical), \ac{BLR} and \ac{AD} (optical - \ac{UV} - X-rays) may also contribute to the observed radiation. The \ac{AD} is directly visible in a few blazars, but is often outshone by the non-thermal jet synchrotron continuum. 
 
 The high-energy \ac{SED} components in the X-ray through gamma-ray energy-bands can be modeled with leptonic or hadronic models \citep{2013ApJ...768...54B}. Both these models provide Compton scattering components in the X-ray through gamma-ray spectrum, which are, however, often sub-dominant in the case of hadronic models. Specifically, they have in common a \ac{SSC} component where electrons Compton up-scatter synchrotron photons previously produced by the same population of electrons.

In the leptonic model, leptons (electrons and possibly positrons) dominate the \ac{EM} radiation. It is possible for protons to be present in the emission region, but protons are assumed not to be accelerated to sufficiently high energies to provide a significant radiative output. The contributions of protons are included when studying the kinetic jet power, which may still be dominated by protons due to their larger rest mass compared to electrons/positrons.  
In the leptonic model the high-energy emission can be produced through both \ac{SSC} and \ac{EC} scattering of low energy seed photons from the \ac{AD}, dust torus and the \ac{BLR} \citep{1985ApJ...298..114M, 1993ApJ...416..458D, 1994ApJ...421..153S}. 

In the hadronic model, electron synchrotron emission dominates at low-frequencies and protons are assumed to be accelerated to sufficiently high energies to radiate appreciably. In the X-ray through gamma-ray regime, radiation is produced by photo-pion interactions yielding \ac{EM} particle cascades (p$+ \gamma \rightarrow \pi^0 +$p or p$+\gamma \rightarrow$ n$+\pi^+$).
In the strong ($\sim 10 - 100$ G) magnetic fields required for hadronic models, ultra-relativistic protons are also efficiently radiating proton-synchrotron radiation \citep{1993A&A...269...67M, 2000NewA....5..377A, 2003APh....18..593M}.

SED modeling leaves many parameter degeneracies, which can be constrained by including polarization information. Spectropolarimetric observations of blazars are particularly useful for this purpose. Such spectropolarimetric observing campaigns are currently being conducted by several groups and at several observatories, including the \ac{SALT}. Ongoing \ac{ToO} spectropolarimetry and spectroscopy observations of blazars are conducted via a \ac{SALT} Large Science Proposal.

In this paper, we discuss and interpret observations of the \ac{FSRQ} \mbox{4C+01.02} (also known as \mbox{PKS B0106+013}; ICRS coord. (ep=J2000): $\alpha=01$h~08m~38.77s, $\delta=+01^\circ~35'~00''.32$ [Optical]; \citealp{2018yCat.1345....0G}), located at a redshift $z=2.1$ \citep{2018A&A...613A..51P}.

This source exhibited its brightest flare ever recorded in gamma rays in 2016 July--August \citep{2016ATel.9232....1V}. Based on this event, we monitored the source at multi-wavelengths for several months, and observed it again in 2017 July--August to uncover the source radiation during a low-state.

The \ac{SED}s of \ac{FSRQ}s are often successfully interpreted in the framework of leptonic models \citep{2012ApJ...752L...4M}. The \ac{FSRQ} \mbox{4C+01.02} has previously been studied by \cite{2011MNRAS.411..901G} and \cite{2017ApJ...851...33P} who determined its \ac{BH} mass as \mbox{$5 \times 10^9 \, \rm M_{\astrosun}$}. The \ac{BH} mass estimate by \cite{2011MNRAS.411..901G}, stemmed from a fit to the optical regime in the \ac{SED} that was strongly dominated by direct \ac{AD} emission, and thus expected to yield low degrees of polarization. However, during flaring states the synchrotron emission is sufficiently dominant to produce a significant degree of total polarization. The transition from low-polarization \ac{AD}-dominated emission to synchrotron-dominated high-polarization emission is accessible to spectropolarimetry and provides an important constraint unavailable with the \ac{SED} alone.  We have detected such a transition in spectropolarimetric observations with \ac{SALT}, motivating a joint fit to the \ac{SED} and the spectropolarimetry.%

In our model, the spectropolarimetry observations are fitted simultaneously with the \ac{SED} to disentangle the spectrum of the synchrotron and \ac{AD} components and thereby, indirectly, constraining the mass of the \ac{BH}. Constraints from spectropolarimetry observations were not considered in most previous blazar models (for the inclusion of spectropolarimetry observations in the modeling of the \ac{FSRQ} \mbox{3C 345}, see \citealp{1986ApJ...305..484S}). 

In this paper, the observations of \mbox{4C+01.02} during its flaring (2016 July--August) and quiescent (2017 July--August) states are described in \mbox{Section~\ref{sec:Observations}}. A simultaneous \ac{SED} and spectropolarimetry model is constructed for blazars in the optical/\ac{UV} regime in \mbox{Section~\ref{sec:ModelSetup}}. The  model is compared with the observations and the results are shown in \mbox{Section~\ref{sec:Results}}. A summary and conclusion of the significance of including spectropolarimetry observations are discussed in \mbox{Section~\ref{sec:SummaryAndConclusions}}. Throughout this paper, we use cosmological parameters $\Omega_{\Lambda}=0.7$, $\Omega_m=0.3$ and $H_0=70 \rm \; km \; s^{-1} \; Mpc^{-1}$. With these parameters, the redshift of $z = 2.1$ corresponds to a luminosity distance of $d_L = 4.952 \times 10^{28} $~cm. 
\section{Observations}\label{sec:Observations}

In this section we describe the observations we conducted in the optical band with the \textit{Las Cumbres Observatory} (LCO) network of telescopes and the \ac{SALT}.  We also describe our analysis of data from the \emph{Swift} X-Ray Telescope (XRT) and the \textit{Large Area Telescope} onboard the \textit{Fermi} Gamma Ray Space Observatory (\textit{Fermi}-LAT) in the MeV-GeV domain.
Radio through \ac{UV} archival data were taken from the NED (\url{http://ned.ipac.caltech.edu/}), WISE (\url{ https://irsa.ipac.caltech.edu}), GALEX (\url{http://galex.stsci.edu/GR6/)}.

\begin{figure*}
  \begin{center}
    \includegraphics[width=\textwidth]{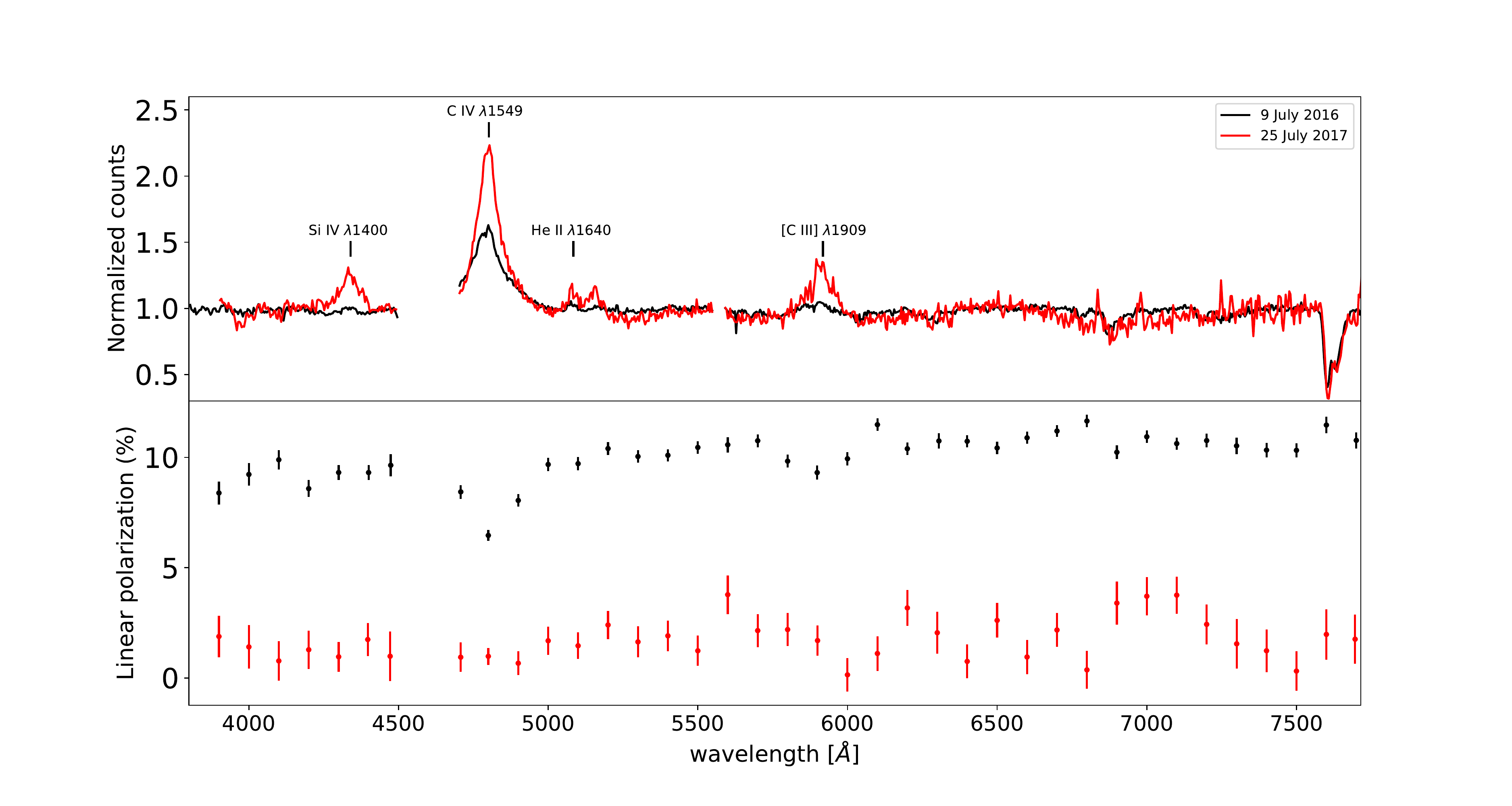}
    \caption{Optical spectropolarimetry data of 4C+01.02 taken on 2016 July 9 (black) and 2017 July 25 (red). \textbf{Top:} Normalized intensity spectra. \textbf{Bottom:} Degree of linear polarization in 100\,\AA{} bins. The gap in the spectrum around 4700\, \AA{} is due to the gap between the RSS CCDs, while the smaller gap around 5570\, \AA{} is a region excluded due to contamination from a sky line.}
    \label{fig:salt_spectrum}
  \end{center}
\end{figure*}
\subsection{Photometry with LCO}

The \emph{Las Cumbres Observatory} (LCO)\footnote{\url{https://lco.global}} was used to obtain photometric observations of 4C+01.02 
in the B, V and R bands during MJD 57602--57605 and MJD 57626--57643, covering parts of the July--August 2016 gamma-ray outburst (see Section~\ref{subsec:LAT}). Observations on 2016 August 2 (MJD 57602.3301--57602.3350) were used to model the optical flux during the flaring state. A set of four observations (B, V, R and I filters) were also taken on 2017 July 28 (MJD 57962.1671--57962.1796) in order to model the optical flux during quiescence. Standard pre-reduction was done with the BANZAI pipeline\footnote{\url{https://github.com/LCOGT/banzai}} 
and the apparent magnitude was calibrated using the SDSS magnitudes of the nearest 10 comparison sources in the field of view, converted to the Johnson-Cousin magnitudes using  \citet{jordi06}. 

\subsection{Optical spectropolarimetry with SALT}

Optical spectropolarimetry observations of 4C+01.02 were carried out on 2016 July 9 (MJD 57578.1354--57578.1638) and on 2017 July 25 (MJD 57959.1062--57959.1347) using the Robert Stobie Spectrograph (RSS) on SALT \citep{SALT06,RSS03a,RSS03b,nordsieck12,2010SPIE.7735E..17B,2016SPIE.9908E..2KP}. Observations were taken in {\sc linear} mode using the pg0300 grating at a grating angle of $5.37^\circ$ with an exposure time of 2400\,s (600\,s per half waveplate angle). The observations covered a wavelength range of $\sim$3200--8500\,\AA{} with a resolving power of $R\sim167$--533. The pre-reduction, wavelength calibration, and polarization measurement were done using the {\sc polsalt} reduction pipeline version 20171226 (\verb!specpolextract_dev 20180524!), based on pySALT v0.5dev.\footnote{\url{https://github.com/saltastro/polsalt}} 
The normalized counts spectrum and degree of polarization are shown in Fig.~\ref{fig:salt_spectrum}.

\begin{table*}
  \caption{Spectral parameters, integral flux and significance of the spectral curvature obtained from the analysis of the \emph{Fermi}-LAT data over the four periods of the 2016 outburst, and the 2017 Quiescent period. We model the spectral shape of 4C+01.02 by a the power-law function (PL --- characterised by the photon index $\Gamma_{PL}$), then by a log-parabola (LP --- characterised by the $\alpha$ and $\beta$ parameters). The integral flux $F$ was determined using the LP model. The $TS_{curv}$ parameters represents the test statistics $(\sim \sigma^2)$ of the spectral curvature of the SED.}
  \begin{center}
    \begin{tabular}{l|c|c|c|c|c}
      \hline
      Period                 & $\Gamma_{PL}$        & $\alpha$        & $\beta$         & $F(0.1-300$~GeV) ($10^{-7}~$ph~cm$^{-2}$~s$^{-1}$) & $TS_{curv}$\\
      \hline
      2016 May 11--May 28 (Pre-flare) & 2.26 $\pm$ 0.06 & 2.04 $\pm$ 0.09 & 0.19 $\pm$ 0.06 & 5.3 $\pm$ 0.4 & 12.9\\
      2016 May 28--Jul 2 (Plateau)   & 2.36 $\pm$ 0.03 & 2.27 $\pm$ 0.04 & 0.11 $\pm$ 0.03 & 9.9 $\pm$ 0.3 & 34.5\\
      2016 Jul 2--Jul 20 (Flare)     & 2.26 $\pm$ 0.03 & 2.11 $\pm$ 0.04 & 0.16 $\pm$ 0.03 & 15.6 $\pm$ 0.6 & 22.4\\
      2016 Jul 20--Aug 15 (Post-flare) & 2.41 $\pm$ 0.03 & 2.32 $\pm$ 0.04 & 0.12 $\pm$ 0.03 & 13.8 $\pm$ 0.4 & 9.4\\
      \hline
      2017 Jul 3 --Aug 2 (Quiescent) & 2.35$\pm$  0.10 & 2.00 $\pm$ 0.15 & 0.28 $\pm$ 0.09 & 0.9 $\pm$ 0.1 & 5.0\\ 
      \hline
    \end{tabular}
    \label{tab:Spec_Params}
  \end{center}
\end{table*}
\begin{figure*}
  \begin{center}
    \includegraphics[width=\textwidth]{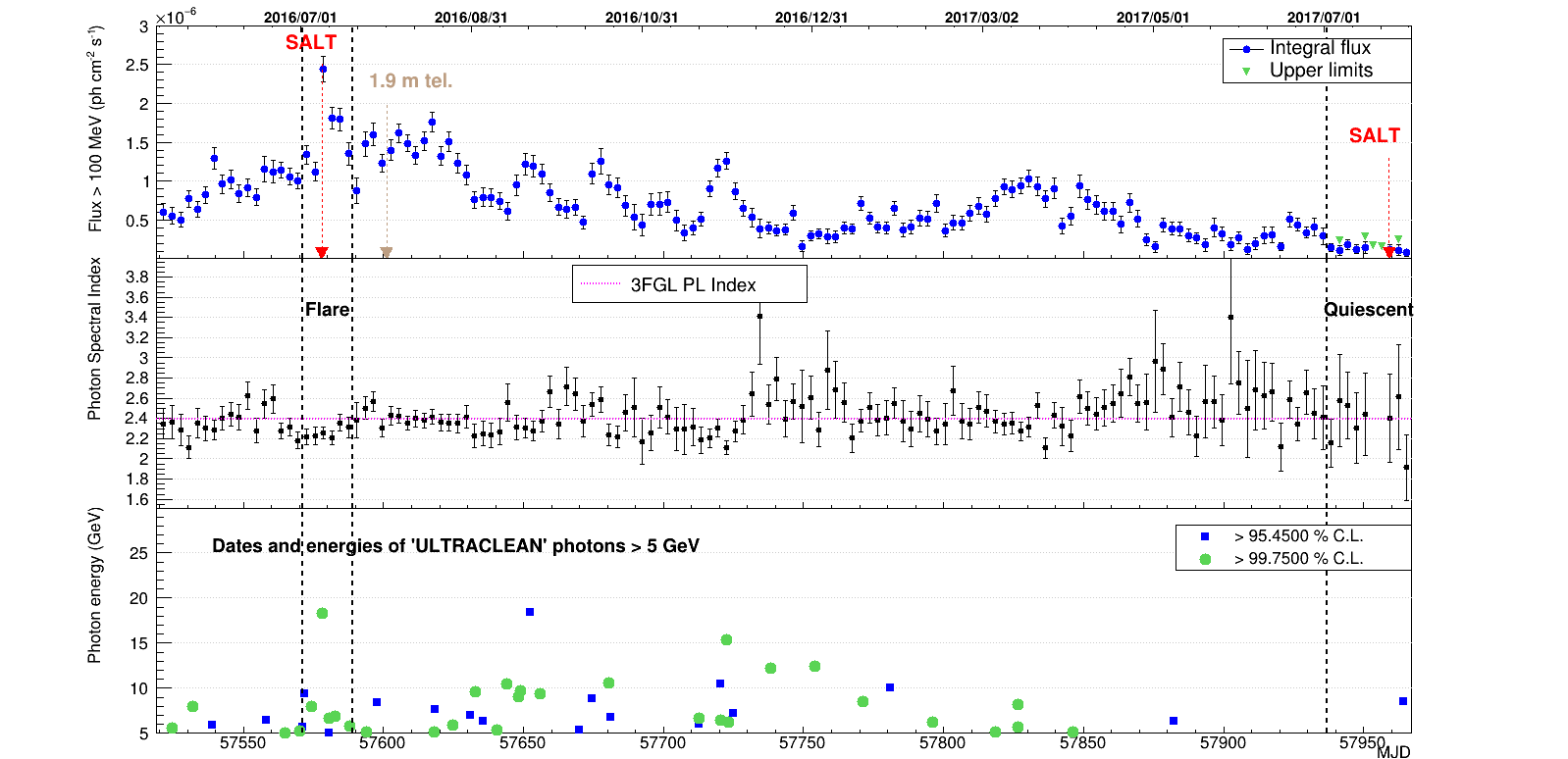}
    \caption{{\bf Top}: \emph{Fermi}-LAT light-curve of 4C+01.02 above 100 MeV between 2016 May 11 and 2017 August 2 in a three-day binning. {\bf Middle}: Corresponding values of the modeled power-law spectral index of 4C+01.02. {\bf Bottom}: energy and date of individual high energy events identified as photons from 4C+01.02 with a high probability. The \emph{Flare} and \emph{Quiescent} periods, corresponding to the \emph{2016 July 2--20} and \emph{2017 July 3--August 2} time ranges respectively, are delimited by black dashed vertical lines.}
    \label{fig:LC_LAT}
  \end{center}
\end{figure*}
The 2016 July 9 data were obtained where 4C+01.02 was close to the maximum of its outburst in that year, during the phase we labeled as ``Flare" or ``Main flare" in Section \ref{subsec:LAT}. On the other hand, the 2017 July 25 observation corresponds to a quiescent episode that lasted for several months in 2017. Since the continuum emission was fainter in 2017, emission lines appear more prominent in the figure. Thanks to the high redshift of the source ($z = 2.1$), prominent ultra-violet emission lines Ly$\alpha$ 1216\,\AA{}, Si~{\sc iv} 1400\,\AA{}, C~{\sc iv} 1549\,\AA{} and C~{\sc iii} 1909\,\AA{} were observed in the 3700--6000 \AA{} range. However, since the Ly$\alpha$ 1216\,\AA{} line was only visible at the extreme blue end of the spectrum where the count rate is low and the spectrum is noisy, it has not been included in Fig.~\ref{fig:salt_spectrum}. 

A spectrum in the 3000--8500 \AA{} range was also taken with the SpUpNIC spectrograph on the SAAO 1.9-m telescope\footnote{\url{https://www.saao.ac.za/astronomers/1-9m/}} on 2016 August 1 (MJD 57601.10) (during the ``post-flare" period,  see Tab.\ref{tab:Spec_Params}) which confirmed the identification of the optical lines mentioned above \citep{HEASA2016_021}.

The degree of linear polarisation is around 10 \% during the \emph{Flare} observation, and between 1 and 2 \% during the \emph{Quiescent} state, which agrees with the prediction that strongly polarized synchrotron emission is released by blazars during outburst episodes.

The unpolarized star GAIA 2538240223562516480 (apparent magnitude G=16.32; distance=491 pc) was observed in the spectrograph's slit of 4C+01.02 on 2017 July 25. The data reduction of this comparison/reference star yields a linear polarisation degree below 1\% for most of the binned data points above 4800~\AA{}. This allows us to estimate a systematic uncertainty on the measurement of the linear polarisation degree.

\begin{figure*}
  \begin{center}
    \includegraphics[width=\textwidth]{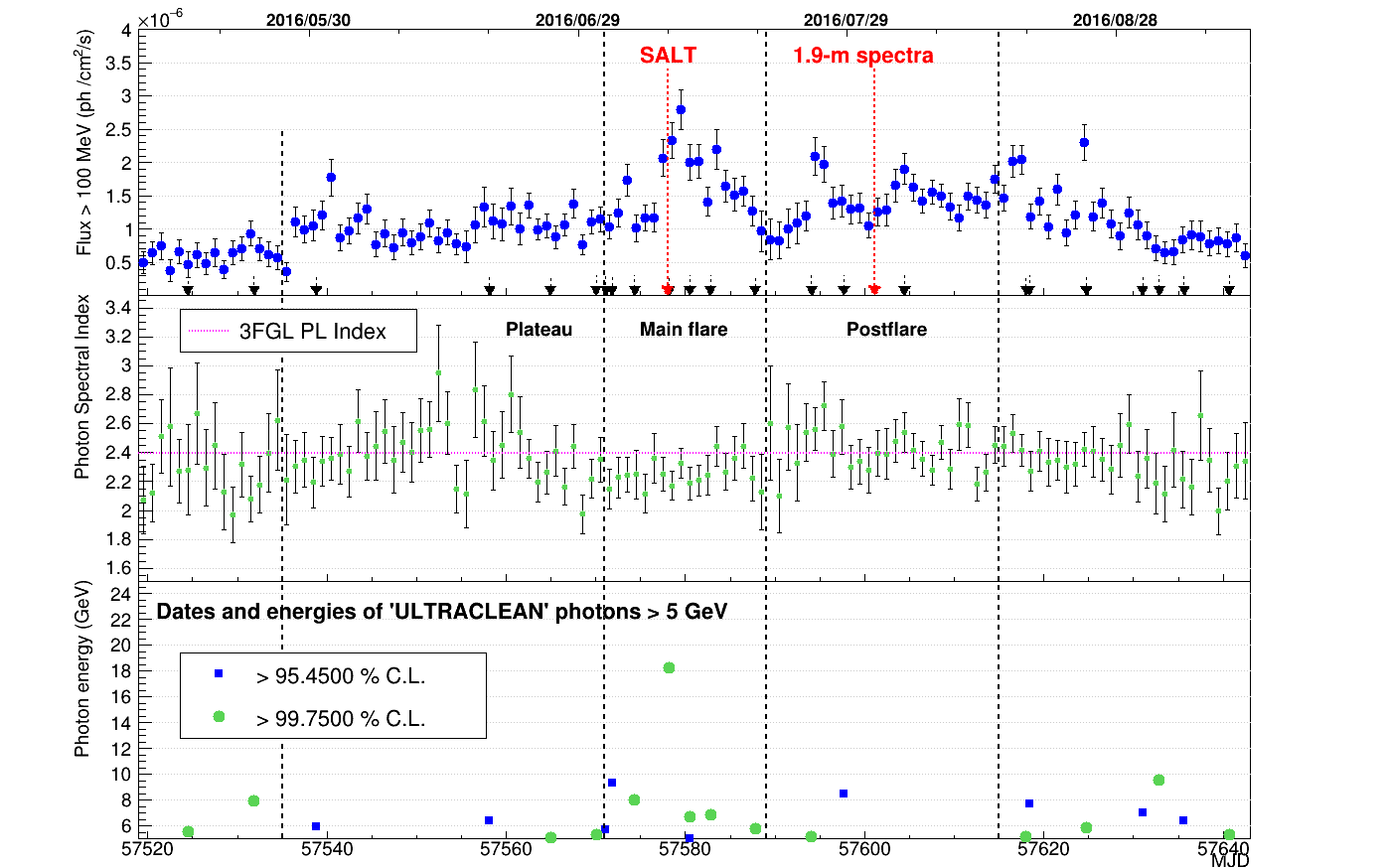}
    \caption{{\bf Top}: \emph{Fermi}-LAT light-curve of 4C+01.02 above 100 MeV between 2016 May 11 and September 12 in a one-day binning. {\bf Middle}: Corresponding values of the modeled power-law spectral index of 4C+01.02. {\bf Bottom}: energy and date of individual high energy events identified as photons from 4C+01.02 with a high probability. The \emph{pre-flare}, \emph{plateau}, \emph{flare} and \emph{post-flare} periods, corresponding to the \emph{2016 May 11--28}, \emph{2016 May 28--July 2}, 2016 July 2--20 and 2016 July 20--August 15 time ranges respectively, are delimited by black dashed vertical lines.}
    \label{fig:LC_LAT_1day}
  \end{center}
\end{figure*}

\subsection{X-ray observations with \emph{Swift}-XRT}

Launched on 2004 November 20, the Neil Gehrels Swift Observatory (\emph{Swift}) \citep{Gehrels:2004aa} is a NASA led space observatory. One of its three instruments, the X-Ray Telescope (XRT) \citep{2005SSRv..120..165B}, is sensitive to soft X-ray photons.

Since no X-ray observation were performed during the 2016 flaring period, we included the \emph{Swift}-XRT SED from \citet{2011MNRAS.411..901G}, built from summed observations from 2007/07/02, 2008/01/10, 2008/02/16 and 2009/08/16, obtained in the 0.2--10 keV range. By considering both the lack of outburst reports and the monitoring of this source from \emph{Fermi}-LAT since August 2008, we can consider that these XRT data represent a moderate quiescent state level of 4C+01.02 and we used them as a lower limit guide in our fit of the broad-band SED of the 2016 flaring period.

\begin{figure*}
  \begin{center}
    \includegraphics[width=\textwidth]{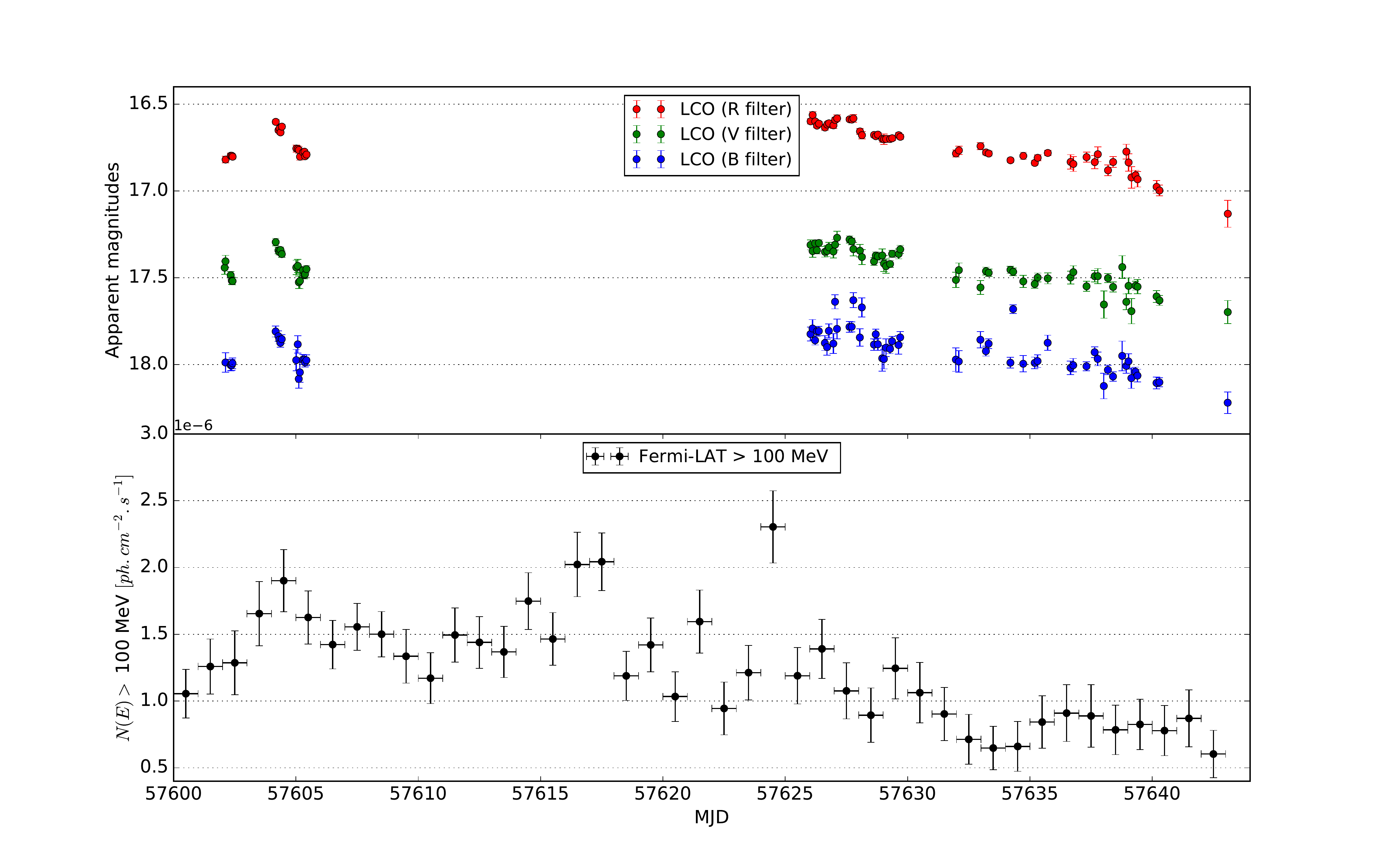}
    \caption{{\bf Top}: LCO light-curves in the R, V and B bands (apparent magnitudes). {\bf Bottom}: \emph{Fermi}-LAT light-curve in a one-day binning.}
    \label{fig:LC_LAT_LCO}
  \end{center}
\end{figure*}

For the quiet state of the source in 2017, we used \textit{Swift}-XRT data taken on \mbox{2017 August 2} (MJD 57967.91), which corresponds to the end of the \emph{Fermi}-LAT \emph{Quiescent} observation period (2017 July 3--August 2).
The cleaned level 3 event files\footnote{\url{http://www.swift.psu.edu/monitoring/source.php?source=PKS0106+01}} generated by \texttt{xrtpipeline}-v.0.13.4 were utilized to generate the image and spectrum within the 0.3--10.0 keV energy range using the \texttt{XSELECT} package from HEASoft v.6.26. The cleaned level 3 event files\footnote{\url{http://www.swift.psu.edu/monitoring/source.php?source=PKS0106+01}} generated by \texttt{xrtpipeline}-v.0.13.4 were utilized to generate the image and spectrum within the 0.3--10.0 keV energy range using the \texttt{XSELECT} package from HEASoft v.6.26.0. First, a circular source region of size 20 pixels and an annular background region of size 50 pixels were chosen to extract the spectrum. The exposure maps created by the \texttt{xrtpipline}-v.0.13.4 were then used to generate the Ancilliary Response File, i.e., the \texttt{arf} file employing the \texttt{xrtmkarf} command.  The Response Matrix File, i.e., the \texttt{rmf} file used in this process was later used for grouping all these spectral files, using the \texttt{grppha} command.

\texttt{Xspec} v.12.10.0c was employed to fit the grouped spectrum with a simple powerlaw model absorbed by Tuebingen-Boulder interstellar medium (ISM) absorption model using the  form: \texttt{tbabs*powerlaw}.

During the fitting procedure, the gas column density between the source and the observer was fixed to the Galactic hydrogen density $N_{\rm H} = 2.42 \times 10^{20} \, \rm cm^{-2}$. This was derived from the LAB survey \citep{2005A&A...440..775K} using the online \texttt{nH} tool\footnote{\url{https://heasarc.gsfc.nasa.gov/cgi-bin/Tools/w3nh/w3nh.pl}}. The goodness of the fit was evaluated by using C-Statistic resulting in $\textrm{C-Stat} =53.52$ for 41 degrees of freedom \citep{1979ApJ...228..939C}. The data that was available was limited to the energy range 0.45--6.0 keV. The final fit yielded an unabsorbed flux of $F_{0.45 -6 \; \rm{keV}} = 1.19 \times 10^{-12} \, \rm erg ~cm^{-2}~s^{-1}$ corresponding to a power law with photon index of $1.43$ $\pm$ $0.23$.

\subsection{Gamma-ray observations with \textit{Fermi}-LAT}  \label{subsec:LAT}

The \emph{Fermi} Gamma-Ray Space Telescope has been orbiting Earth since June 2008. It operates in survey mode most of the time, covering the whole sky every $3 \,$h (corresponding to two orbits), thanks to its large (\mbox{$\sim2.4$ sr}) field of view. Following the 2018 March 16 solar panel drive anomaly, at some sun angles, some sources may receive less exposure on ~1-week timescales.\footnote{\url{https://fermi.gsfc.nasa.gov/ssc/observations/types/post\_anomaly/}} This allows a regular monitoring of sources on the whole sky. Its main instrument, the Large Area Telescope (LAT), is sensitive to photons from $\sim$20~MeV to several hundreds of GeV \citep{2009ApJ...697.1071A}.

\begin{figure*}[!ht]
\centering
\subfigure{\includegraphics[width=.45\textwidth]{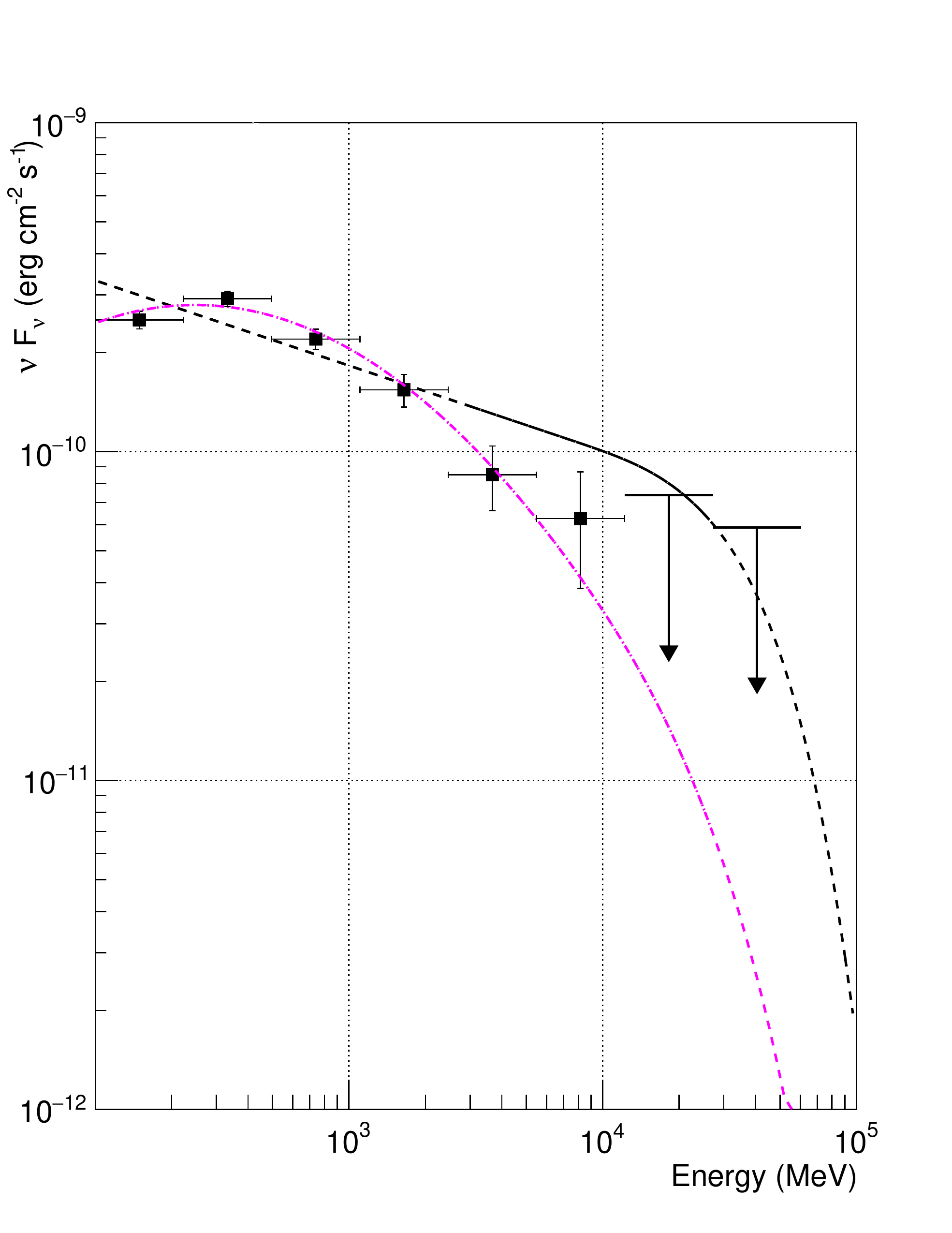}}
\subfigure{\includegraphics[width=.45\textwidth]{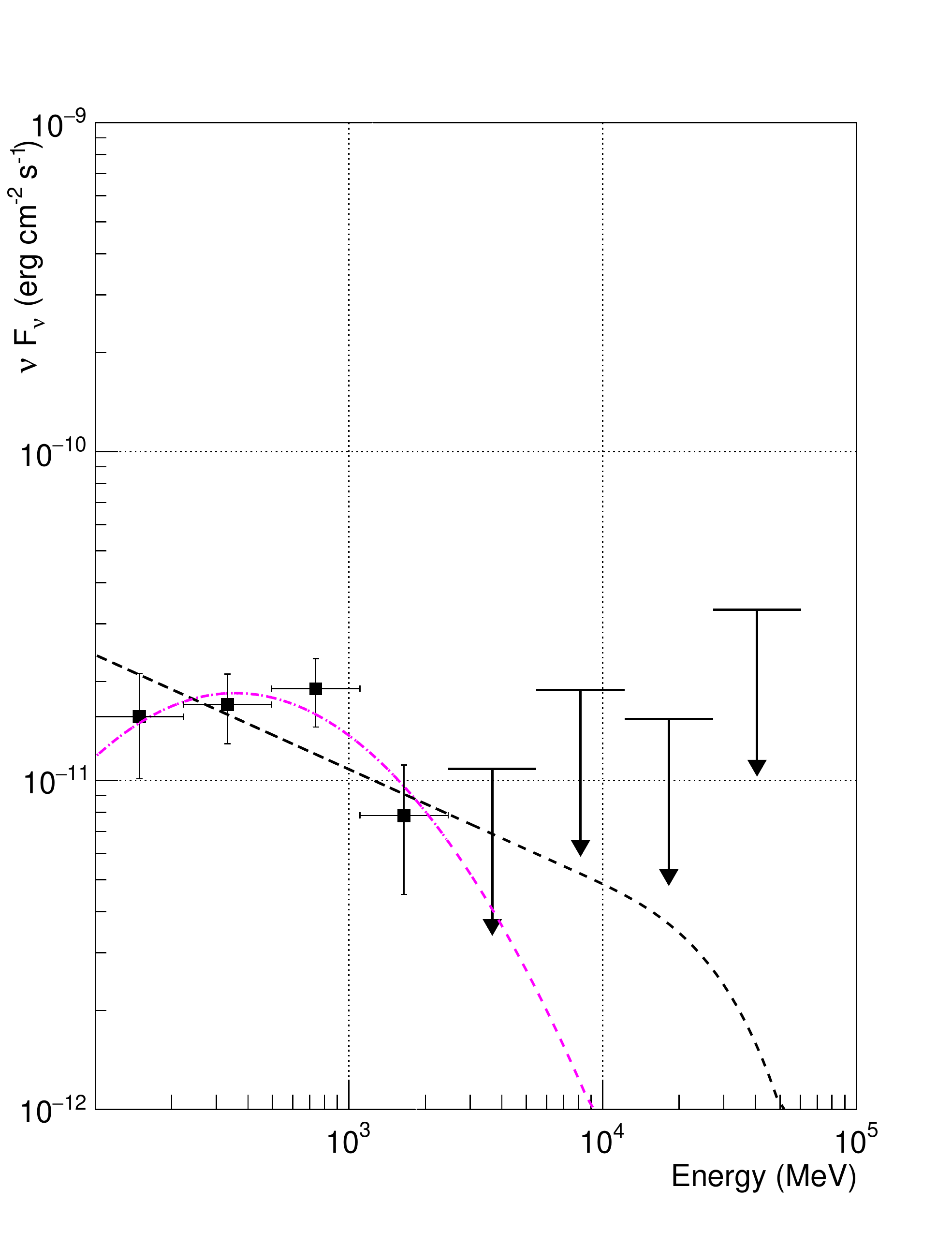}}
\caption{\label{fig:SED} \emph{Fermi}-LAT \ac{SED} of \mbox{4C+01.02} during the ``flare" (left) and ``quiescent" (right) states. Data points are fitted by a PL (black dashed) and an LP (magenta dashed) function. The absorption by the extragalactic background light was modeled by the $e^{-\tau_{\gamma\gamma}(E)}$ factor applied to both functions and corresponding to the redshift $z = 2.1$, using the model developed by \citet{2010ApJ...712..238F}.}
\end{figure*} 

We present the analysis of \emph{Fermi}-LAT data from 2016 May till 2017 October, 
in the 100 MeV--300 GeV range. We used the Pass8 (R2) dataset \citep{p8art}
, and the \emph{Fermi} Science Tools version v10r0p5.\footnote{\url{http://fermi.gsfc.nasa.gov/ssc/data/analysis/}} We performed both time domain and spectral analysis by running the unbinned likelihood algorithm (gtlike/pyLikelihood Science Tool) with the following standard analysis cuts applied to point source analysis: radius of the \emph{Region of interest} (ROI)=$15^\circ$; Source region: a $10^\circ$ annulus surronding the ROI; {\sc source} class; event type = 3; zenith angle $< 90^\circ$; {\texttt DATA\_QUAL=1, LAT\_CONFIG=1}; Diffuse emission templates: \texttt{gll\_iem\_v06.fits} (Galactic) and \texttt{iso\_P8R2\_SOURCE\_V6\_v06.txt} (isotropic). 
The spectrum of the source of interest was successively modeled by the two following functions: a log-parabola (LP -- with its standard parameters $\alpha$ and $\beta$) and a single power law (PL -- with photon index $\Gamma_{PL}$), except while running the likelihood algorithm in narrow time or energy bins. Depending on the data set and analysis cuts, between 8 and 16 parameters defining the spectral shapes of the brightest point sources of the ROI and the spectral index of the two diffuse templates were kept free in the likelihood analysis.

We present in Figures~\ref{fig:LC_LAT} and \ref{fig:LC_LAT_1day} the three-day and daily light-curves, respectively, of 4C+01.02 (top panel), the time evolution of the PL spectral index (middle panel) and the dates and energy of high energy photons above 5 GeV, identified as gamma-rays and as emitted by 4C+01.02 with a probability $>95.45 \% $ to originate from the target (bottom panel). According to the flux level and variability pattern, we defined four episodes that we referred to as \emph{pre-flare}, \emph{plateau}, \emph{(main) flare} and \emph{post-flare}, successively. In Table~\ref{tab:Spec_Params} we present the spectral parameters that we obtained in our analysis of each of these four episodes, and the 2017 subset of the quiescent state, using successively the PL and LP functions to model the spectral shape of 4C+01.02. The \emph{plateau--flare--post-flare} pattern was previously observed for FSRQ 3C~454.3 during several of its outbursts (eg, \citet{2011ApJ...733L..26A, 2016ApJ...830..162B} and references therein).


In Figure~\ref{fig:LC_LAT_1day}, the epochs of SALT and SAAO observations are highlighted with red arrows. The variation of the spectral photon index $\Gamma_{PL}$ (middle panel) suggests a hardening of the SED when the source is brighter --- during the main flare. This is a common feature reported for bright FSRQs during outbursts. An unusual feature is that the pre-flare period (corresponding to a relatively quiescent state of the source) seems to be also characterised by a hardening of the spectrum, as also reported in Table~\ref{tab:Spec_Params}.\\

This table also lists the test statistic that quantifies the presence of spectral curvature in the gamma-ray spectrum of the source. This was done by computing  $TS_{\rm curv}$ as follows:
\begin{equation}
TS_{\rm curv} = -2 \times [ ln (\mathcal{L}_{LP} ) - ln (\mathcal{L}_{PL} )],
\end{equation}
where $ln(\mathcal{L}_{LP})$ and $ln(\mathcal{L}_{PL})$ represent the natural logarithm of the maximum likelihood obtained with the LP and PL models, respectively. 
The preferred model is LP.

We show in Figure~\ref{fig:LC_LAT_LCO} a subset of the \emph{Fermi}-LAT light-curve of Figure~\ref{fig:LC_LAT_1day}, along with the LCO light-curve obtained during the 2016 observation campaigns. LCO data presented here include dereddening. However, the LCO observations were not continuous and most of the contemporaneous \emph{Fermi}-LAT/LCO monitoring was undertaken after the main flare when the variability of the source was not significant in gamma-rays. This prevented us from quantifying any time lag between the different energy bands, though the visual inspection reveals that the peak positions around MJD 57604 appears without significant shift between the B, V, R and LAT bands. We also observe a flux decrease after MJD 57626 for all the four bands. This observation suggests a single-zone origin of the optical and gamma-ray emissions.

Spectral analysis is presented for both the flare and quiescent periods (Figure \ref{fig:SED}). Fits were performed using both the PL and LP functions with all their parameters fixed to the values obtained from the unbinned likelihood analysis shown in Table~\ref{tab:Spec_Params}. (SED data points were obtained using PL models for each point source in the ROI and source region.)

The estimated systematic uncertainty in the effective area is 5\% in the 100~MeV--100~GeV range. The energy resolution ($\Delta E/E$, at 68\% containment) is 20\% at 100 MeV, and between 6 and 10\% over the 1--500 GeV range.\footnote{\url{http://fermi.gsfc.nasa.gov/ssc/data/analysis/LAT\_caveats.html}} \footnote{\url{http://www.slac.stanford.edu/exp/glast/groups/canda/lat\_\\Performance.htm}}

\color{black} 

\section{Model Setup}\label{sec:ModelSetup}

A model is constructed for blazars that simultaneously fits the low-energy (microwave through \ac{UV}) component of the \ac{SED} and the degree of polarization as a function of wavelength in the optical-\ac{UV} regime. The simultaneous \ac{SED} and polarization fit consists of the synchrotron, \ac{AD} and emission lines from the \ac{BLR} flux components. 
 
The host galaxy and dust torus emissions are considered negligible in the strongly jet-dominated IR - optical spectrum, but can be included in the model in future work. The synchrotron flux is calculated from a broken power-law, exponential cut-off electron distribution 
\begin{equation}
\label{eq:electron_distribution}
N_e(\gamma) = n_{\rm 0} \left \lbrace \begin{array}{lcl}
(\frac{\gamma}{\gamma_{\rm b}})^{-p_1} \cdot e^{-\gamma_{\rm b}/ \gamma_{\rm c}} \textrm{for } \gamma_{\rm min} \le \gamma \le \gamma_{\rm b} \\
(\frac{\gamma}{\gamma_{\rm b}})^{-p_2} \cdot e^{-\gamma/\gamma_{\rm c}} \textrm{    for } \gamma_b < \gamma < \gamma_{\rm max}
\end{array} \right. ,
\end{equation}
where $n_0$ is a normalization factor determining the total number of non-thermal electrons, $\gamma_{\rm b}$ and $\gamma_{\rm c}$ the characteristic break and cut-off energies in the co-moving frame of the emission region and $p_1$ and $p_2$ the electron spectral indices. The emission region is pervaded by a magnetic field B and moves along the jet with a bulk Lorentz factor $\Gamma$. To reduce the number of free parameters, we assume a viewing angle of $\theta_{\rm obs} = 1/\Gamma$ so that the Doppler factor $\delta = (\Gamma [ 1 - \beta_{\Gamma} \cos\theta_{\rm obs}])^{-1} = \Gamma$. Where the emission region becomes optically thick (at optical depth $\tau_{SSA}= R_{em} \cdot \alpha_{SSA}> 1$), the synchrotron emission is self-absorbed. Here, $R_{em}$ is the emission region radius. The absorption coefficient $\alpha_{SSA}$ is calculated following Eq. (6.50) of \cite{1986rpa..book.....R}. The \ac{SSA} effect leads to a steepening of the spectrum below a critical \ac{SSA} frequency where \mbox{$\tau_{SSA} = 1$}.  

The \cite{1973A&A....24..337S} \ac{AD} model is used for calculating the \ac{AD} \ac{SED} component. This model assumes a geometrically thin \ac{AD} around a non-rotating \ac{BH} for which the innermost stable orbit is $R_{\rm in} = 3 \, R_{\rm S}$, with $R_{\rm S}$ the Schwarzschild radius. The outer boundary of the \ac{AD} is assumed to be at radius $R_{\rm out} = 10^3 \, R_{\rm S}$. Our results are insensitive to the exact choice of $R_{\rm out}$ as the cold disk at those radii no longer contributes significantly to the overall spectrum. At the inner disk radius, the maximum disk temperature is defined as:

\begin{equation}
\label{eq:T_max}
T^{\rm AD, \: max} \propto l^{1/4}M_{\rm BH}^{-1/4},
\end{equation}
with \mbox{$l = \frac{\dot{M}_{\rm BH}}{\dot{M}_{\rm Edd}}=\frac{L_{\rm AD}}{L_{\rm Edd}}$} being the Eddington ratio and the accretion rate $\dot{M}_{\rm BH} = L_{\rm AD}/(\epsilon c^2)$. The efficiency of converting potential energy into radiation is assumed as $\epsilon = 1/12$. The \ac{AD} \ac{SED} component peaks at a frequency which is related to this maximum temperature through
\begin{equation}
\label{eq:nu_peak}
\nu^{\rm AD, \: peak}=2.8\cdot kT^{\rm AD, \; max}/h.
\end{equation} 
The parameters determining $\nu^{\rm AD, \: peak}$ are $M_{\rm BH}$ (increasing $M_{\rm BH}$ yields lower  $\nu^{\rm AD, \: peak}$), $\epsilon$ (increasing $\epsilon$ increases $\nu^{\rm AD, \: peak}$ and yields higher $\nu F^{\rm AD}_{\nu}$) and $\dot{M}_{\rm BH}$ (increasing $\dot{M}_{\rm BH}$ increases $\nu^{\rm AD, \: peak}$ and $\nu F^{\rm AD}_{\nu}$) \citep{Calderone}.

Synchrotron polarization is calculated as 
\begin{equation}
\Pi^{\rm syn}_{\omega} =F_{B} \cdot \frac{\int N_e(\gamma)x(\gamma)K_{2/3}(x(\gamma))d\gamma}{\int N_e(\gamma)x(\gamma)\int^{\infty}_{x(\gamma)}K_{5/3}(x(\xi))d \xi d\gamma},\label{eq:synchrotron_polarisation}
\end{equation}
where $F_B$ is the factor characterizing the ordering of the magnetic field and $x(\gamma) = \omega/\omega_c(\gamma)$, with $\omega_c(\gamma)$ the critical frequency \citep{1959ApJ...130..241W}. This parameterization of the magnetic field ordering has previously been used in \cite{1986ApJ...305..484S} and \cite{2013ApJ...774...18Z}. The Bessel functions $K_{2/3}$ and $K_{5/3}$ are computed with the python inbuilt Bessel function \textit{besselk()} from the \textit{mpmath} package. 

For a pure power-law electron spectrum with index $p$ (corresponding to a synchrotron radiation spectral index $\alpha = [p - 1]/2$), the degree of synchrotron polarization can be \mbox{$\Pi^{\rm syn}_{\rm max} = \frac{p+1}{p+7/3} = \frac{\alpha+1}{\alpha+5/3}$} \citep{2011hea..book.....L}. For spectral indices in the range \mbox{$2<p<3$}, the maximum degree of synchrotron polarization is \mbox{$69\% < \Pi^{\rm syn}_{\rm max} < 75 \%$}. For a broken power-law distribution, the synchrotron polarization is still approximately described by the above identities for frequencies sufficiently far away from (between) the spectral breaks/cut-offs. For a gradually steepening synchrotron spectrum (due to a broken / cut-off electron distribution), the synchrotron polarization increases towards higher frequencies, corresponding to an increasing value of $\alpha$. 

The total optical/UV flux is the sum of the polarized synchrotron and unpolarized AD and emission line fluxes. Hence, the total degree of polarization is given by
\begin{equation}
\Pi^{\rm tot}_{\omega} = \frac{\Pi^{\rm syn}_{\omega} \cdot F^{\rm syn}_{\omega}}{F^{\rm syn}_{\omega} + F^{\rm AD}_{\omega} + F^{\rm lines}_{\omega}}.
\label{eq:TotalPolarization}
\end{equation}

In the model code, the unpolarized emission lines are approximated by Gaussian functions. The emission line fluxes are calculated relative to each other and independently of the continuum flux \citep{Francisetal1991}.

A $\chi^2$ minimization technique was employed to determine the best-fit parameters characterizing the non-thermal electron distribution, $F_B$, \ac{AD} luminosity and BH mass. 

The broadband \ac{SED} is subsequently obtained by employing the steady-state leptonic blazar model of \cite{2013ApJ...768...54B} with the non-thermal electron spectrum and magnetic field obtained from the low-frequency \ac{SED} and spectropolarimetry fit described above. The \ac{SED} and spectropolarimetry fit determines the density of radiating electrons. The characteristics of the external radiation field are adjusted, to obtain a fit to the high-energy (X-ray through gamma-ray) \ac{SED}. The BLR target photon field is modelled as a thermal radiation field that is isotropic in the AGN rest frame with a characteristic temperature such that the resulting EC spectrum is a good representation of a BLR radiation field. For a comparison between using a detailed, line-dominated BLR radiation field and a thermal BLR, see \cite{2013ApJ...768...54B}.

\section{Results and Discussion}\label{sec:Results}
The model was fitted to contemporaneous LCO optical photometry and SALT spectropolarimetry data, complemented by archival radio through UV data of \mbox{4C+01.02} in its flaring state from 2016 and quiescent state from 2017. The fit was conducted over the \ac{SALT} spectropolarimetry observations
 in the \mbox{$3.9 \times 10^{14}$ Hz} to \mbox{$7.5 \times 10^{14}$ Hz} range. The results of this \ac{SED} and spectropolarimetry fitting are discussed in \mbox{Section~\ref{sec:SimulModel}}. A fit of the broadband SED, including the high-energy (Compton) components produced with the code of \cite{2013ApJ...768...54B} is discussed in \mbox{Section~\ref{sec:BroadBand}}. We discuss and compare our results to those obtained in previous work in \mbox{Section~\ref{sec:Comparison}} and to the \ac{BH} mass estimate based on the C~{\sc iv} line width and the continuum luminosity in \mbox{Section~\ref{sec:BHmass}}.

The parameter results obtained by the model presented in this paper are indicated by a superscript $M$, and appended by the superscripts $f$ and $q$ to represent the flaring and quiescent states, respectively.

\subsection{Simultaneous SED and Spectropolarimetry Fit Results}\label{sec:SimulModel}
\begin{figure*}[!ht]
\centering
\includegraphics[scale=0.64]{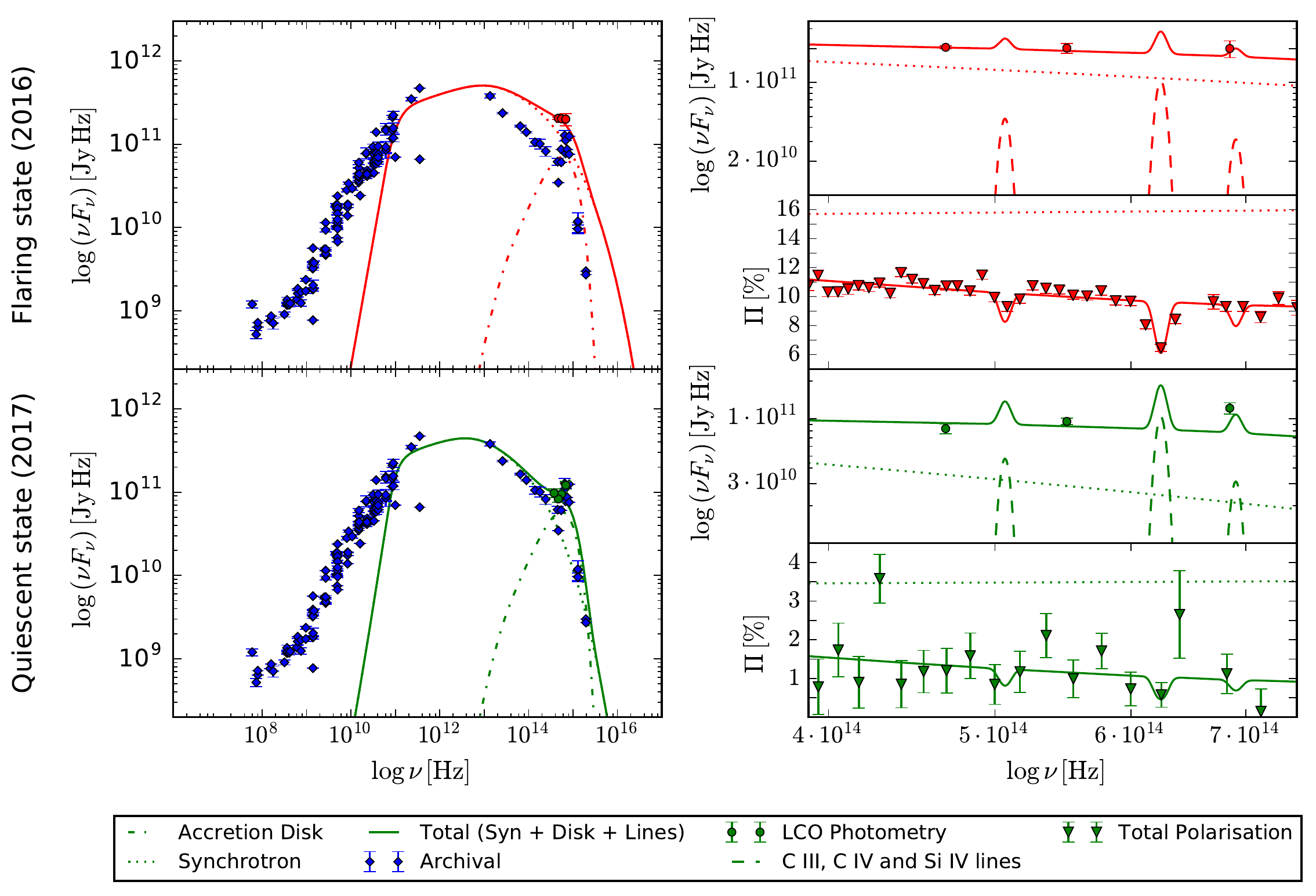}
\caption{\label{fig:SimulModel}The simultaneous low-energy \ac{SED} and spectropolarimetry model (lines) fitted to the observations (data points) of \mbox{4C+01.02} in its flaring state of 2016 (top panels, red) and quiescent state of 2017 (bottom panels, green). The \ac{SED}s are plotted in the left panels and the \ac{SED}s and degree of polarization components in the optical/\ac{UV} frequency range are plotted in the right panels. Model components and observational data are shown in the legend.}
\end{figure*}

The results of the simultaneous \ac{SED} and degree of polarization fit to the observations of 4C+01.02 in its flaring and quiescent states in 2016 and 2017, respectively, are plotted in Figure~\ref{fig:SimulModel} with corresponding fit parameters and quantities derived from the fit parameters shown in Table~\ref{table:SimulModel}. The low-energy \ac{SED} components are shown in the left panels where the total flux (contributed by the synchrotron, \ac{AD} and BLR lines flux components) is fitted through the optical LCO photometry data in the B, V, R filters for the flaring state and the data in the B, V, R, I filters for the quiescent state (the I filter data point is not visible in the plots to the right as it is not in the polarization data regime). The right panels show the \ac{SED} and degree of polarization components in the optical/\ac{UV} regime. The decrease in the total degree of polarization (due to the unpolarized \ac{AD} emission diluting the synchrotron emission towards higher optical frequencies) constrains the \ac{AD} component, thereby, disentangling the \ac{AD} flux component and the synchrotron flux component. By constraining the \ac{AD} component during the quiescent state the \ac{BH} mass of \mbox{4C+01.02} is determined as $\MBH \, \rm M_{\astrosun}$. This obtained \ac{BH} mass was also adopted in the flaring state. When archival \ac{UV} observations are not taken into account during the quiescent state model fit, a lower \ac{BH} mass of $4 \times 10^8 \, \rm M_{\astrosun}$ is obtained \citep{FermiSymProc}.

The spectropolarimetry fits clearly show the decrease of the polarization at the frequencies of the unpolarized \mbox{C {\sc iii}}, \mbox{C {\sc iv}} and \mbox{Si {\sc iv}} emission lines. It is assumed that the emission line fluxes do not change significantly from the flaring to the quiescent state, as the emission line flux in the quiescent state is poorly constrained. 

The radio data are not fitted due to the steepening of the one-zone synchrotron spectrum below the \ac{SSA} critical frequency, at $\nu_{SSA}\sim 2 \times 10^{11} \, \rm Hz$ for both flaring and quiescent states. The higher optical flux in the flaring state is achieved through a larger synchrotron emission component. Note that the electron spectrum normalization constant $n_0$ represents the differential number of electrons at $\gamma_b$, where $\gamma_b$ is significantly larger in the flaring state fit compared to the quiescent state fit. Thus, the fit parameters imply a larger number of high-energy electrons in the flaring state. 

The \ac{AD} luminosity is obtained through the model fit as $L^{M, \, f}_{\rm AD} = \Flareld \, \rm erg ~ s^{-1}$ and $L^{M, \, q}_{\rm AD} = \Quiesld \, \rm erg ~ s^{-1}$ for the flaring and quiescent states, respectively. The corresponding maximum temperatures $T^{M, \, f}_{\rm max}= \Flaretm \, \rm K$ and $T^{M, \, q}_{\rm max}= \Quiestm \, \rm K$ of the \ac{AD} near the innermost stable circular orbit yields a peak frequency of $\nu^{M, \, f}_{T, \: \rm max}=\Flarevmio \, \rm Hz$ and $\nu^{M, \, q}_{T, \: \rm max}=\Quiesvmio \, \rm Hz$ for the flaring and quiescent states, respectively, which is dependent on the \ac{AD} luminosity and \ac{BH} mass through \mbox{Equation~\ref{eq:T_max}}. 

Both states necessitated a partially ordered magnetic field with $F_B <1$. The higher ordering of the magnetic field in the flaring state $F_B^{M, \:f}=\Flarefb$ indicates 4.7-fold increase of the magnetic field ordering parameter compared to the quiescent state, where $F_B^{M, \: q}=\Quiesfb$. 

The $\chi^2$ values per degree of freedom $n$ for the degree of total polarization fit are $(\chi^2/n)^{M, \:f}_{\rm pol} = \Flarechi$ and $(\chi^2/n)^{M, \:q}_{\rm pol} = \Quieschi$. These values appear acceptable when keeping in mind that additional features in the spectrum have not been included in the model fit such as additional faint emission lines and possibly absorption lines, and a dusty torus which may yield a small contribution in diluting the synchrotron polarization. In this first exploration of combined SED and spectropolarimetry fitting, we fit only the jet synchrotron continuum, the direct \ac{AD} emission, and the most prominent emission lines.

\begin{deluxetable*}{cCCCCCC}
\tablecaption{\label{table:SimulModel}Parameters and calculated values obtained from the simultaneous \ac{SED} and polarization model fit for \mbox{4C+01.02} in its flaring state of 2016 (second column) and quiescent state of 2017 (third column). The electron distribution in the emission frame has characteristic Lorentz factors $\gamma$ and the characteristic synchrotron radiation profile frequencies are defined in the observer frame. }
\tablehead{ \colhead{Parameters} & \colhead{Flaring state} & \colhead{Quiescent state} }
\decimals
\startdata
\multicolumn{3}{c}{Input Parameters}\\
\hline 
\textrm{Magnetic field B [G] at emission region height}  & $\Flareb $ & $\Quiesb$ \\[0.2em]
\textrm{Bulk Lorentz factor} $\Gamma$ & $\FlareGamma$ & $\QuiesGamma$ \\[0.2em]
\textrm{Emission region radius} $R_{em}$ \textrm{[cm]}  & $\Flarebr$ & $\Quiesbr$  \\[0.2em]
\hline 
\multicolumn{3}{c}{Parameters Obtained with Fit}\\
\hline 
\textrm{Normalization factor} $n_{\rm 0}$ & $\Flarenz$ & $\Quiesnz$ \\[0.2em]
\textrm{Minimum gamma} $\gamma_{\rm min}$ & $\Flaregoe$ & $\Quiesgoe$ \\[0.2em]
\textrm{Gamma break} $\gamma_{\rm b}$ & $\Flaregbe$ & $\Quiesgbe$\\[0.2em]
\textrm{Critical gamma} $\gamma_{\rm c}$ & $\Flaregce$  & $\Quiesgce$\\[0.2em]
\textrm{Electron spectral index} $p_1$ & $\Flarepo$ & $\Quiespo$\\[0.2em]
\textrm{Electron spectral index} $p_2$ & $\Flarept $ & $\Quiespt $\\[0.2em]
\textrm{Ordering of magnetic field} $ F_{B}$ & $\Flarefb$ & $\Quiesfb$\\[0.2em]
\textrm{Disk luminosity} $ L_{\rm AD}$ [$\rm erg~s^{-1}$] & $\Flareld$ & $\Quiesld$ \\[0.2em]
\textrm{BH mass} [$\rm M_{\astrosun}$] & $\MBH$ & $\MBH$\\[0.2em]
\textrm{C~{\sc iv} line flux} $h_2 \, \rm [Jy~Hz]$ & $\Flareht$ & - \\[0.2em]
\hline 
\multicolumn{3}{c}{Calculated Values}\\
\hline 
\textrm{Minimum frequency} $\nu_{\rm min}$ \textrm{[Hz]} & $\Flarevmo$ & $\Quiesvmo$\\[0.2em]
\textrm{Break frequency}  $\nu_b$ \textrm{[Hz]}& $\Flarevbo$ & $\Quiesvbo$\\[0.2em]
\textrm{Critical frequency} $\nu_c$ \textrm{[Hz]}  & $\Flarevco$  & $\Quiesvco$\\[0.2em]
\textrm{Photon spectral index} $\alpha_1$ & $\Flareao$ & $\Quiesao$\\[0.2em]
\textrm{Photon spectral index} $\alpha_2$ & $\Flareat $ & $\Quiesat $\\[0.2em]
\textrm{Maximum synchrotron polarization} $\Pi^{\rm syn}_{\rm max}(p_1)$ & $\Flareso $ & $\Quiesso $\\[0.2em]
\textrm{Maximum synchrotron polarization} $\Pi^{\rm syn}_{\rm max}(p_2)$ & $\Flarest $ & $\Quiesst $\\[0.2em]
\textrm{Maximum disk temperature } $T_{AD, \;\rm max}$ \textrm{[K]} & $\Flaretm$ & $\Quiestm$ \\[0.2em]
$\nu_{T, \: \rm max}$ \textrm{[Hz]} & \Flarevmio & \Quiesvmio \\[0.2em]
\textrm{Goodness of fit} $(\chi^2/n)_{\rm pol}$ & $\Flarechi$ & $\Quieschi$ \\[0.2em]
\textrm{C~{\sc iii} line flux} $h_1 \, \rm [Jy~Hz]$ & $\Flareho$ & $\Quiesho$\\[0.2em]
\textrm{C~{\sc iv} line flux} $h_2 \, \rm [Jy~Hz]$ & - & $\Quiesht$\\[0.2em]
\textrm{Si~{\sc iv} line flux} $h_3 \, \rm [Jy~Hz]$ & $\Flarehth$ & $\Quieshth$\\[0.2em]
\enddata
\end{deluxetable*}

\subsection{Broad-Band Spectral Energy Distribution}\label{sec:BroadBand}
\begin{figure*}[!ht]
\centering
\includegraphics[scale=0.5]{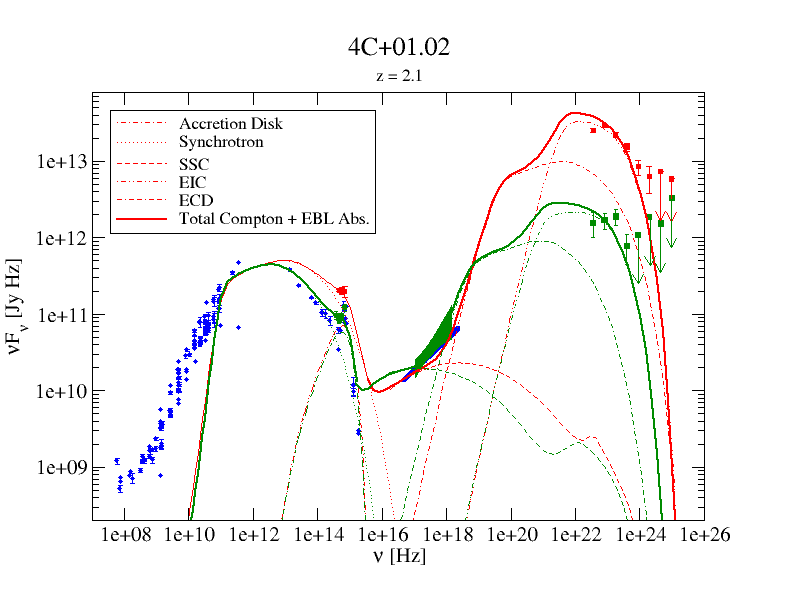}
\caption{\label{fig:BroadbandSED}Broad-band \ac{SED} of \mbox{4C+01.02} during its flaring (red) and quiescent (green) states by implementation of the low-energy parameters obtained during the model fit (given in Table~\ref{table:SimulModel}) into the code of \cite{2013ApJ...768...54B} to produce the X-ray and gamma-ray components for a leptonic model. Data from observations are as follow: archival data points (blue), LCO photometry data points for the flaring (red) and quiescent (green) states, \textit{Swift}-XRT data points for the flaring (blue) and quiescent (green) states, and \textit{Fermi}-LAT data points for the flaring (red) and quiescent (green) states.\\
}
\end{figure*}

The electron distribution obtained during the model fit (\mbox{Table~\ref{table:SimulModel}}) was subsequently used as parameters in the \cite{2013ApJ...768...54B} code to model the X-ray and gamma-ray emission resulting from \ac{SSC} and \ac{EC} scattering. The resulting broadband \ac{SED} fits are plotted in \mbox{Figure~\ref{fig:BroadbandSED}} with parameters as listed in \mbox{Table~\ref{table:4C+01.02HEParameters}}. 

The emission region radius is kept the same between the flaring and quiescent states, in order to reduce the number of varying parameters. 

 Our fits suggest a decreasing accretion rate from the flaring to quiescent state. The fits further require that the emission region is placed slightly further down the jet in the quiescent state compared to the flaring state, which could lead to a suppressed \ac{EC} (\ac{AD}) component, and could also be consistent with the lower energy density of the \ac{BLR} radiation field ($u^{M, \: q}=\Quiesu \, \rm erg~cm^{-3}$ for the quiescent state compared to the flaring state's $u^{M, \: f}=\Flareu \, \rm erg~cm^{-3}$).

The dominant \ac{EC} (\ac{BLR}) flux component during the flaring state suggests that the emission region is within the \ac{BLR}, providing an approximately isotropic energy density. During the quiescent state the \ac{EC} (\ac{BLR}) flux component is suppressed, suggesting that the emission region is located at or beyond the outer boundary of the \ac{BLR}. 

\begin{deluxetable*}{cCC}
\tablecaption{\label{table:4C+01.02HEParameters}Parameters obtained by implementation of the low-energy parameters constrained with the simultaneous \ac{SED} and polarization model (Table~\ref{table:SimulModel}) in the \cite{2013ApJ...768...54B} code to predict the X-ray and gamma-ray components of the leptonic model in \mbox{Figure~\ref{fig:BroadbandSED}}.}
\tablehead{ \colhead{Parameters} & \colhead{Flaring state} & \colhead{Quiescent state} }
\decimals
\startdata
\multicolumn{2}{c}{Input Parameters}\\
\hline
\textrm{Kinetic luminosity in jet electrons [$\rm erg~s^{-1}$]} & $\Flarekelum$ & $\Quieskelum$\\[0.2em]
\textrm{Emission region height} $z_{\rm 0}$ \textrm{[pc]}  & $\Flareemh$ & $\Quiesemh$ \\[0.2em]
\textrm{Observing angle} $\theta_{\rm obs}=1/\Gamma$ \textrm{[$^\circ$]} & $\Flareobsa$ & $\Quiesobsa$ \\[0.2em]
\textrm{External radiation field energy density} $u$ [$\rm erg~cm^{-3}$] &   $\Flareu$  & $ \Quiesu$ \\[0.2em]
\textrm{External radiation field black body temperature [K]}      & $\Flarebbt$ & $\Quiesbbt$\\[0.2em]
\hline 
\multicolumn{2}{c}{Derived Parameters}\\
\hline 
 $L_B (jet)$ [$\rm erg~s^{-1}$]& $\Flarelb$ & $\Quieslb$\\[0.2em]
$ L_B/L_e$ & $\Flarelble$ & $\Quieslble$ \\[0.2em]
 $dt_{\rm var, \: min}$ & $\Flaredtvars  \, \rm s$ & $\Quiesdtvars \, \rm s$ \\[0.2em] 
$ $   & $= \Flaredtvarh \, \rm h$ & $= \Quiesdtvarh \, \rm h$
 \\ 
\enddata
\end{deluxetable*}

\subsection{Comparison to Previous Work}\label{sec:Comparison}
\begin{figure*}[!ht]
\centering
\includegraphics[scale=0.64]{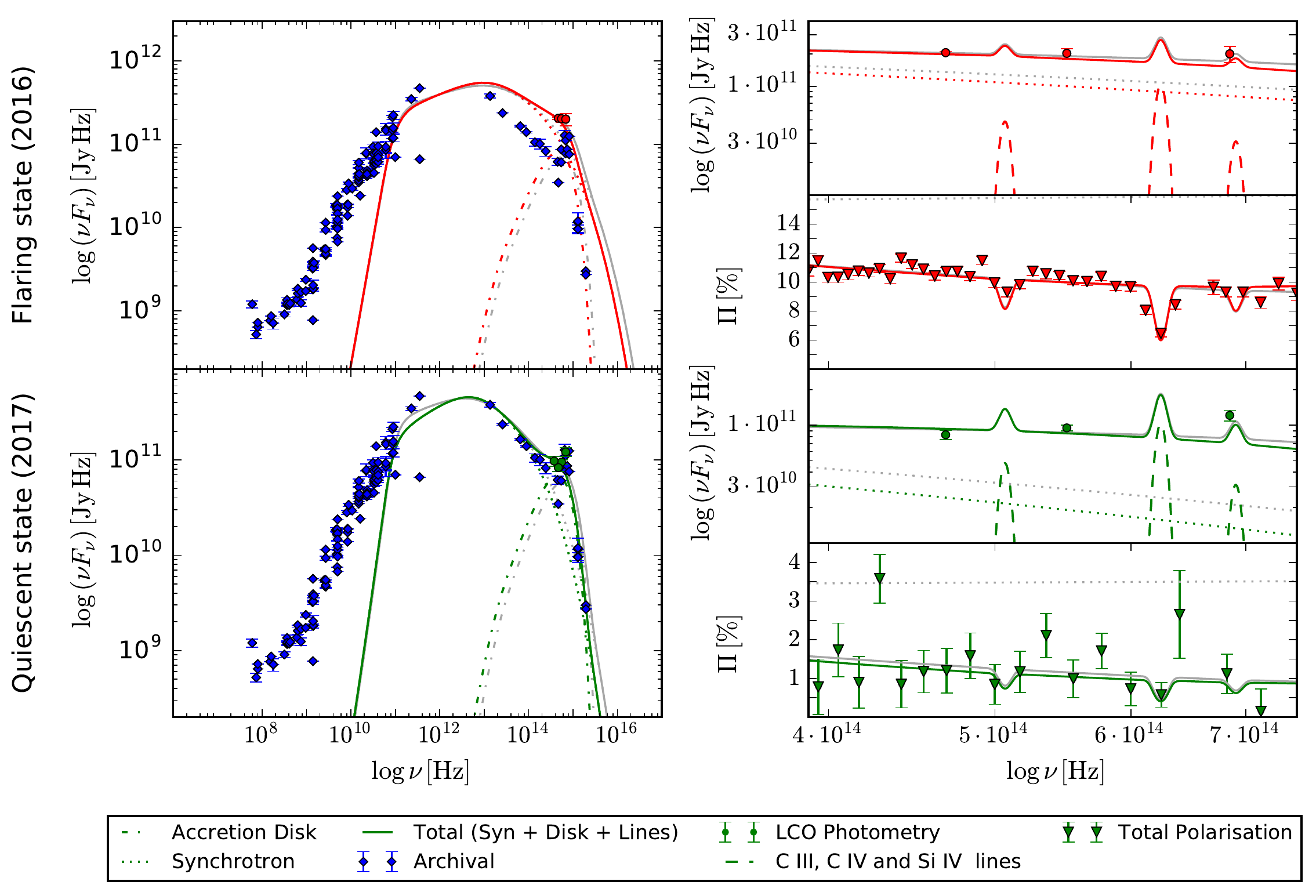}
\caption{\label{fig:PeakShift}Implementation of the \ac{BH} mass obtained for \mbox{4C+01.02} as $\GhiMBH \,\rm  M_{\astrosun}$ by \cite{2011MNRAS.411..901G} and \cite{2017ApJ...851...33P} in our simultaneous \ac{SED} and spectropolarimetry model. Parameters obtained during this model fit and calculated values are given in \mbox{Table~\ref{tab:ComparisonPS}}. The model fit obtaining a \ac{BH} mass of $\MBH \, \rm M_{\astrosun}$ are shown in gray. }
\end{figure*}

Previous works modeling \mbox{4C+01.02} were of \cite{2011MNRAS.411..901G} and \cite{2017ApJ...851...33P} considering model fits to \ac{SED} observations, without including any polarization degree information. \cite{2011MNRAS.411..901G} considered a model fit to \ac{SED} observations with optical emission being strongly dominated by an \ac{AD}. \cite{2011MNRAS.411..901G} and \cite{2017ApJ...851...33P} inferred $M_{\rm BH} \GhiMBH \rm ~M_{\astrosun}$. In order to test whether flaring and/or quiescent state spectropolarimetry data can safely exclude a larger BH mass, we fixed the BH mass in our fit routine to their obtained \ac{BH} mass of $\GhiMBH \rm ~ M_{\astrosun}$ and attempted \ac{SED} and spectropolarimetry fits. Parameters and inferred quantities are listed in \mbox{Table~\ref{tab:ComparisonPS}} and the resulting low-frequency \ac{SED} and spectropolarimetry fits are shown in \mbox{Figure~\ref{fig:PeakShift}}.

During the flaring state of \mbox{4C+01.02}, the IR-optical-\ac{UV} spectrum is expected to be dominated by the jet synchrotron emission and not the \ac{AD} emission. The quiescent state may be \ac{AD} dominated with a low ordered magnetic field in the jet emission region.


Our model, using our best-fit \ac{BH} mass and that of \cite{2011MNRAS.411..901G} and \cite{2017ApJ...851...33P} (their results are indicated by superscript $C$) produces \ac{AD} flux components peaking at \mbox{$\nu_{T, \: \rm max}^{M, \: f} = \Flarevmio \, \rm Hz$} and $\nu_{T, \: \rm max}^{M, \: q} = \Quiesvmio \, \rm Hz$, and \mbox{$\nu_{T, \: \rm max}^{C, \: f} = \FlarevmioPS \, \rm Hz$} and {$ \nu_{T, \: \rm max}^{C, \: q} = \QuiesvmioPS \, \rm Hz$}, respectively. The degree of total polarization, thereby, decreases at lower frequencies, as demonstrated by Figure~\ref{fig:PeakShift}, second and forth right panels.

The reduced chi-square $\chi^2/n$ values for the fits to $n$ amount of spectropolarimetry data points, are $(\chi^2/n)^{C, \: f}_{\rm pol} = \FlarechiPS$ and $(\chi^2/n)^{C, \: q}_{\rm pol} = \QuieschiPS$, compared to $(\chi^2/n)^{M, \: f}_{\rm pol} = \Flarechi$ and $(\chi^2/n)^{M, \: q}_{\rm pol} = \Quieschi$. 
This demonstrates that spectropolarimetry data disentangles the synchrotron and AD \ac{SED} flux components (left in Figure~\ref{fig:PeakShift}) and determines the ordering of the magnetic field. The \ac{BH} mass obtained by \cite{2011MNRAS.411..901G} and \cite{2017ApJ...851...33P} cannot be excluded.

\begin{deluxetable*}{cCCCCCC}
\tablecaption{\label{tab:ComparisonPS}Parameters obtained and values calculated by implementing the \ac{BH} mass of $\GhiMBH \, \rm M_{\astrosun}$ obtained for \mbox{4C+01.02} by \cite{2011MNRAS.411..901G} and \cite{2017ApJ...851...33P}. These results correspond to \mbox{Figure~\ref{fig:PeakShift}} where the \ac{AD} components was shifted during the flaring and quiescent states to enable a fit.}
\tablehead{ \colhead{Parameters} & \colhead{Flaring state} & \colhead{Quiescent state} }
\decimals
\startdata
\multicolumn{3}{c}{Input Parameters}\\
\hline 
\textrm{Magnetic field B [G] at emission region height}  & $\FlarePSb$ &  $\QuiesPSb$ \\[0.2em]
\textrm{Bulk Lorentz factor} $\Gamma$ & $\FlarePSGamma $ & $\QuiesPSGamma $ \\[0.2em]
\textrm{Emission region radius} $R_{em}$ \textrm{[cm]}  & $\Flarebr$ & $\Quiesbr$  \\[0.2em]
\hline
\multicolumn{3}{c}{Parameters Obtained with Fit}\\
\hline 
\textrm{Normalization factor }$n_{\rm 0}$ & $\FlarenzPS$ & $\QuiesnzPS$ \\[0.2em]
\textrm{Minimum gamma} $\gamma_{\rm min}$ & $\FlaregoePS$ & $\QuiesgoePS$ \\[0.2em]
\textrm{Gamma break }$\gamma_{\rm b}$ & $\FlaregbePS$ & $\QuiesgbePS$\\[0.2em]
\textrm{Critical gamma }$\gamma_{\rm c}$ & $\FlaregcePS$  & $\QuiesgcePS$\\[0.2em]
\textrm{Electron spectral index }$p_1$ & $\FlarepoPS$ & $\QuiespoPS$\\[0.2em]
\textrm{Electron spectral index }$p_2$ & $\FlareptPS $ & $\QuiesptPS $\\[0.2em]
\textrm{Ordering of magnetic field }$ F_{B}$ & $\FlarefbPS$ & $\QuiesfbPS$\\[0.2em]
\textrm{Disk luminosity }$ L_{\rm AD}$ [$\rm erg~s^{-1}$] & $\FlareldPS $  & $\QuiesldPS $ \\[0.2em]
\textrm{C~{\sc iv} line flux} $h_2 \, \rm [Jy~Hz]$ & $\FlarehtPS$ & - \\[0.2em]
\hline
\multicolumn{3}{c}{Calculated Values}\\
\hline 
\textrm{Minimum frequency }$\nu_{\rm min}$ \textrm{[Hz]} & $\FlarevmoPS$ & $\QuiesvmoPS$\\[0.2em]
\textrm{Break frequency }$\nu_{\rm b}$ \textrm{[Hz]} & $\FlarevboPS$ & $\QuiesvboPS$\\[0.2em]
\textrm{Critical frequency }$\nu_{\rm c}$ \textrm{[Hz]}  & $\FlarevcoPS$  & $\QuiesvcoPS$\\[0.2em]
\textrm{Photon spectral index }$\alpha_1$ & $\FlareaoPS$ & $\QuiesaoPS$\\[0.2em]
\textrm{Photon spectral index }$\alpha_2$ & $\FlareatPS $ & $\QuiesatPS $\\[0.2em]
\textrm{Maximum synchrotron polarization }$\Pi^{\rm syn}_{\rm max}(p_1)$ & $\FlaresoPS $ & $\QuiessoPS $\\[0.2em]
\textrm{Maximum synchrotron polarization }$\Pi^{\rm syn}_{\rm max}(p_2)$ & $\FlarestPS $ & $\QuiesstPS $\\[0.2em]
\textrm{Maximum disk temperature }$T_{AD, \;\rm max}$ \textrm{[K]} & $\FlaretmPS$ & $\QuiestmPS$ \\[0.2em]
$\nu_{T, \: \rm max}$ \textrm{[Hz]} & $\FlarevmioPS$ & $\QuiesvmioPS$ \\[0.2em]
\textrm{Goodness of fit }$(\chi^2/n)_{pol}$ & $\FlarechiPS$ & $\QuieschiPS$ \\[0.2em]
\textrm{C~{\sc iii} line flux} $h_1 \, \rm [Jy~Hz]$ & $\FlarehoPS$ & $\QuieshoPS$\\[0.2em]
\textrm{C~{\sc iv} line flux} $h_2 \, \rm [Jy~Hz]$ & - & $\QuieshtPS$\\[0.2em]
\textrm{Si~{\sc iv} line flux} $h_3 \, \rm [Jy~Hz]$ & $\FlarehthPS$ & $\QuieshthPS$\\[0.2em]
\enddata
\end{deluxetable*}

\subsection{\ac{BH} mass estimate based on C~{\sc iv} line width and continuum luminosity}
\label{sec:BHmass}

We here provide an independent estimate of the \ac{BH} mass, following the C~{\sc iv}-based \ac{BH} mass estimator as recently extended by \cite{Park:2017rqa}. Their Eq. (4) allows a \ac{BH} mass estimate based on the full-width at half max (FWHM) of the C~{\sc iv} $\lambda$1549 emission line (in units of km/s), and the \ac{AD} continuum luminosity at a rest-frame wavelength of 1350~\AA. The latter corresponds to an observed frequency of $\nu^{\rm cont}_{\rm obs} = 7.2 \times 10^{14}$~Hz. From our SED fitting, we find an \ac{AD} continuum $\nu F_{\nu} = \lambda F_{\lambda}$ flux of $4.2 \times 10^{-13}$~erg~cm$^{-2}$~s$^{-1}$ at that frequency, corresponding to a rest-frame luminosity of $\lambda L_{\lambda} = 4 \pi d_L^2 \, \nu F_{\nu} = 1.3 \times 10^{46}$~erg~s$^{-1}$, where we used a luminosity distance of $d_L = 4.95 \times 10^{28}$~cm. 

Fitting a Lorentzian profile to the C~{\sc iv} emission line in our \ac{SALT} spectra, we obtain a FWHM of $\Delta\lambda_{2016} = 109 \, $\AA{} for the observation of 9 July 2016, and $\Delta\lambda_{2017} = 90 \, $\AA{} for the observation of 25 July 2017. We therefore use $\Delta\lambda_{\rm ave} = 100 \,$\AA{}  as a representative average value, corresponding to a velocity of FWHM(C~{\sc iv}) = 6250~km~s$^{-1}$. 

Plugging these values into Eq. (4) of \cite{Park:2017rqa}, we find $\log(M_{\rm BH}/\rm M_{\odot}) = 8.9^{+0.54}_{-0.53}$. In the evaluation of this expression, the systematic error of the $M_{\rm BH}$ -- $L_{1350}$ -- FWHM(C~{\sc iv}) relation greatly dominates over the measurement uncertainties, as the latter are small and only enter logarithmically. We therefore only account for the systematic errors, which are quoted as 1-$\sigma$ uncertainties. The logarithm above then corresponds to
\begin{equation}
M_{\rm BH} = (7.7 \times 10^8)^{+2.2 \times 10^9}_{-5.4 \times 10^8} \, \rm M_{\odot}.
\label{BHCIV}
\end{equation}

The BH mass of $\MBH \, \rm M_{\odot}$ as found through our SED and spectropolarimetry fitting is consistent with the upper limit, while the mass of $\sim 5 \times 10^9 \, \rm M_{\odot}$ used by \cite{2011MNRAS.411..901G} and \cite{2017ApJ...851...33P} is further outside the 1-$\sigma$ error interval. If one interpreted the entire observed flux at $7.2 \times 10^{14}$~Hz during the quiescent state as accretion disk flux, the best estimate would increase to $\sim 1.2 \times 10^9 \, \rm M_{\odot}$, with the upper limit of the 1-$\sigma$ error range increasing to $4.2 \times 10^9 \, \rm M_{\odot}$. 

Thus, we conclude that this independent black-hole mass estimate favours the value of $M_{\rm BH} \MBH \, \rm M_{\odot}$ from our SED and spectropolarimetry fitting.

\section{Summary and Conclusions}\label{sec:SummaryAndConclusions}
In this paper we describe the development of a code to simultaneously fit the low-frequency \ac{SED} and spectropolarimetry data of blazars. The model was applied to the blazar \mbox{4C+01.02}, comparing contemporaneous optical LCO photometry and \ac{SALT} spectropolarimetry data during a flaring and a quiescent state in 2016 and 2017, respectively. The model assumes an unpolarized AD diluting the degree of electron synchrotron polarization towards higher frequencies and by disentangling the synchrotron and AD components, the AD component and the BH mass is constrained. With our model fits, we could determine the \ac{BH} mass of \mbox{4C+01.02} as $\MBH \, \rm M_{\astrosun}$.

The magnetic field ordering determines the synchrotron polarization degree and can be determined with spectropolarimetry observations. This parameter will also determine the degree of high-energy polarization due to synchrotron self-Compton scattering in the X-ray and gamma-ray regimes.

The highly ordered magnetic field $F_B^{M, \: f}= \Flarefb$ during the flaring state, compared to the quiescent state's $F_B^{M, \: q}= \Quiesfb$, indicates a localized emission region possibly representing the passing of a shock. Shocks enhance, compress and order the magnetic fields in the jet and thereby higher synchrotron and \ac{SSC} flux components are obtained from the jet. However, higher \ac{EC} (\ac{AD}) and \ac{EC} (\ac{BLR}) flux components are not dependent on  magnetic field ordering. These components might be enhanced because particles are accelerated more efficiently. A higher/lower synchrotron flux component means that there are more/less available photons to produce a higher/lower \ac{SSC} flux component in the flaring/quiescent state.

The less ordered magnetic field in the quiescent state suggests the presence of more tangled magnetic fields where magnetic field reconnection can take place and magnetic field orientations in different directions cancel each other out. These lead to lower degree of synchrotron polarization.

A fit to the low-frequency (synchrotron + \ac{AD} + \ac{BLR})
 emission components alone can not independently constrain electron energies, magnetic field, and the size of the emission region. Additional constraints on those parameters result from a fit to the entire \ac{SED}, which was subsequently performed using the code of \cite{2013ApJ...768...54B}. The resulting fits suggest that \ac{SSC}, \ac{EC} (\ac{AD}) and \ac{EC} (\ac{BLR}) are all contributing significantly to the X-ray through gamma-ray emission of \mbox{4C+01.02}, and that the emission region is likely to be located further out along the jet in the quiescent state, compared to the flaring state. 

The \ac{BH} mass of $\MBH \, \rm M_{\astrosun}$, obtained with our model fits, is inconsistent with \cite{2011MNRAS.411..901G} and \cite{2017ApJ...851...33P} who obtained the \ac{BH} mass for this source as $\GhiMBH \, \rm M_{\astrosun}$ by fitting the \ac{SED}. Our \ac{BH} mass estimate is consistent with the upper limit of the C~{\sc iv}-based \ac{BH} estimation following the method developed by \cite{Park:2017rqa}. According to a simultaneous fit to the \ac{SED} and spectropolarimetry data, the \ac{BH} mass obtained by \cite{2011MNRAS.411..901G} and \cite{2017ApJ...851...33P} cannot be ruled out.

Both our own model and the models of \cite{2011MNRAS.411..901G} and \cite{2017ApJ...851...33P} considered a \cite{1973A&A....24..337S} \ac{AD} model assuming a non-rotating \ac{BH}. However, if the \ac{AD} is considered as truncated at the innermost stable orbit and the \ac{BH} is considered to rotate retrograde/prograde to the \ac{AD}, the disk moves farther out/closer to the \ac{BH}. The maximum disk temperature then decreases/increases giving the impression of a larger/smaller \ac{BH} mass. Since both models used the same \ac{AD} prescription based on a non-rotating \ac{BH}, \ac{BH} rotation can not be invoked to resolve the discrepancy in the inferred \ac{BH} mass values. The \ac{BH} mass inferred by \cite{2011MNRAS.411..901G} and \cite{2017ApJ...851...33P} results in an \ac{AD} component at lower frequencies in the \ac{EM} spectrum. If we take into account that in their model \cite{2017ApJ...851...33P} used $\epsilon \sim 1/10$ compared to our $\epsilon \sim 1/12$, their \ac{AD} component moves to lower $\nu_{T, \; \rm max}$ and lower fluxes when considering lower $\epsilon \sim 1/12$,  since $\nu_{T, \; \rm max} \propto \epsilon^{1/4}$. This will, therefore, not provide a  shift of their \ac{AD} component to higher frequencies and does not explain the discrepancy \citep{Calderone}.

In future work, we will extend our model to include high-energy (Compton) emission in the same fitting routine, including predictions for X-ray and gamma-ray polarization, and include potential hadronic high-energy emission components, similar to the work of \cite{2013ApJ...774...18Z} and \cite{2018ApJ...863...98P}. Inclusion of the \ac{IC} components will enable further constraints on the emission region parameters and dominant emission mechanisms. Predictions will be made for the Future Imaging X-ray Polarimetry Explorer (IXPE) \citep{2016ResPh...6.1179W} space-craft that is scheduled for launch in 2021\footnote{\url{https://ixpe.msfc.nasa.gov}} and the All-sky Medium Energy Gamma-ray Observatory (AMEGO)\footnote{\url{https://asd.gsfc.nasa.gov/amego/}} for gamma-ray polarimetry.

\ac{SALT} \ac{ToO} observations and co-ordinated multi-wavelength observations are continuing so that more sources (including BL Lacs) will be available for combined \ac{SED}s and spectropolarimetry fitting.

In the case of 4C+01.02 considered here, the host galaxy and a putative dusty torus do not contribute significantly to the observed \ac{SED}. However, for the purpose of fitting other sources, these contributions will be included in our model as additional unpolarized emission components in future work. 

POLLUX, a \ac{UV} polarimeter aboard the \ac{LUVOIR} has been proposed as a Concept Study to the 2020 Decadal Survey which could provide \ac{UV} polarization data to further constrain the \ac{AD} components in blazar \ac{SED}s towards \ac{UV} frequencies \citep{2018SPIE10699E..3BB, 2018sf2a.conf...71M}.

\section*{Acknowledgements}
This research has been made possible through a Large Science Program; the \ac{SALT} \ac{ToO} program ``Observing the Transient Universe". Spectropolarimetry observations reported in this paper were obtained with the \ac{SALT} under program 2016-2-LSP-001 (PI: D.~A.~H. Buckley). This work makes use of observations from the \ac{LCOGT} network under a dedicated transient program (PI: B. van Soelen). Radio through \ac{UV} archival data were taken from NED, WISE and GALEX and the \ac{Swift}-\ac{XRT} data were used.

The \textit{Fermi} LAT Collaboration acknowledges generous ongoing support from a number of agencies and institutes that have supported both the development and the operation of the LAT as well as scientific data analysis. These include the National Aeronautics and Space Administration and the Department of Energy in the United States, the Commissariat \`a l'Energie Atomique and the Centre National de la Recherche Scientifique / Institut National de Physique Nucl\'eaire et de Physique des Particules in France, the Agenzia Spaziale Italiana and the Istituto Nazionale di Fisica Nucleare in Italy, the Ministry of Education, Culture, Sports, Science and Technology (MEXT), High Energy Accelerator Research Organization (KEK) and Japan Aerospace Exploration Agency (JAXA) in Japan, and the K.~A.~Wallenberg Foundation, the Swedish Research Council and the Swedish National Space Board in Sweden.

Additional support for science analysis during the operations phase is gratefully acknowledged from the Istituto Nazionale di Astrofisica in Italy and the Centre National d'\'Etudes Spatiales in France. This work performed in part under DOE Contract DE-AC02-76SF00515.

H.M. Schutte acknowledges support from the \ac{NRF}\footnote{Any opinion, finding and conclusion or recommendation expressed in this material is that of the authors, and the \ac{NRF} does not accept any liability in this regard.} and Centre for Space Research (CSR) in South Africa. 

The work of M. B\"{o}ttcher is supported through the South African Research Chair Initiative (SARChI) of the \ac{NRF} and the Department of Science and Innovation of South Africa, under SARChI Chair grant number 64789. 

B. van Soelen and J.P. Marais are supported by the \ac{NRF} of South Africa (respective grant numbers: 116300, 100259). 

A. Kaur and A.D. Falcone gratefully acknowledge support from NASA grants 80NSSC20K1526 and NNX16AR77G.

We thank Enrico J. Kotze, Ken Nordsieck, Dani\`el Groenewald, Steve Crawford and Steve Potter for providing tools, training and precious advises for the reduction of SALT-RSS spectropolarimetry data sets (and for the use of \texttt{reducepoldataGUI.py v1.3} developed by Enrico Kotze). 

We also thank Vaidehi S. Paliya and Justin Finke for their meticulous reading of the draft and their valuable comments as internal referee. And we thank Sara Buson, Deirdre Horan and Josefa Becerra, Philippe Bruel and Matthew Kerr for their final comments on the draft, on behalf of the \textit{Fermi}-LAT Collaboration.



\begin{acronym}
 \acro{AD}{Accretion Disk}
 \acro{AGN}{Active Galactic Nuclei}
 \acro{BH}{black hole}
 \acro{BLR}{Broad Line Region}
 \acro{EC}{External Compton}
 \acro{EBL Abs.}{Extragalactic Background Light Absorption}
 \acro{ECD}{External Compton (Disc)}
 \acro{EIC}{External Inverse Compton (BLR)}
 \acro{EM}{electromagnetic}
 \acro{Fermi-LAT}{\textit{Fermi} Large Area Telescope}
 \acro{FSRQ}{Flat Spectrum Radio Quasar}
 \acro{GPE}{Gravitational Potential Energy}
 \acro{HBL}{High-frequency peaked BL Lac objects}
 \acro{H.E.S.S.}{\textit{High Energy Stereoscopic System}}
 \acro{IC}{Inverse Compton}
 \acro{IR}{Infrared}
 \acro{KE}{kinetic energy}
 \acro{LAT}{Large Area Telescope}
 \acro{LBL}{Low-frequency peaked BL Lac objects}
 \acro{LCO}{\textit{Las Cumbres Observatory}}
 \acro{LCOGT}{\textit{Las Cumbres Observatory Global Telescope}}
 \acro{LOS}{Line of Sight}
 \acro{LUVOIR}{\textit{Large UV/Optical/Infrared Surveyor}}
 \acro{MAGIC}{\textit{Major Atmospheric Gamma Imaging Cherenkov}}
 \acro{NLR}{Narrow Line Region}
 \acro{NRF}{National Research Foundation}
 \acro{QSO}{Quasi Stellar Object}
 \acro{RSS}{Robert Stobie Spectrograph}
 \acro{SAAO}{\textit{South African Astronomical Observatory}}
 \acro{SALT}{\textit{Southern African Large Telescope}}
 \acro{SED}{Spectral Energy Distribution}
 \acro{SMBH}{Super Massive Black Hole}
 \acro{SS}{Shakura Sunyaev}
 \acro{SSA}{Synchrotron Self-Absorption}
 \acro{SSC}{Synchrotron Self-Compton}
 \acro{Swift}{\textit{Neil Gehrels Swift Observatory}}
 \acro{ToO}{Target of Opportunity}
 \acro{UFS}{University of the Freestate}
 \acro{UV}{ultraviolet}
 \acro{VERITAS}{Very Energetic Radiation Imaging Telescope Array System}
 \acro{XRT}{X-ray Telescope }
\end{acronym}

\software{BANZAI (\cite{2018SPIE10707E..0KM}, \url{https://github.com/LCOGT/banzai}), pySALT (v0.50dev, \cite{2010SPIE.7737E..25C}, \url{https://pysalt.salt.ac.za/}), POLSALT (\cite{polsalt}, \url{https://github.com/saltastro/polsalt}), HEAsoft (v6.26, \cite{2014ascl.soft08004N}, \url{https://heasarc.gsfc.nasa.gov/lheasoft/}), XSPEC (v12.10.0c; \cite{1996ASPC..101...17A}, \url{https://heasarc.gsfc.nasa.gov/xanadu/xspec/}), {\it Fermi} Science Tools (v10r0p5, \url{https://fermi.gsfc.nasa.gov/ssc/data/analysis/software/v10r0p5.html}).}


\bibliographystyle{aasjournal.bst}
\bibliography{references.bib}
\end{document}